\documentclass[twocolumn]{aastex7}

\usepackage{amsmath}
\usepackage{overpic}

\shorttitle{Cold Outflows in BAL Quasars}
\shortauthors{Zhu et al.}

\begin{document}

\title{A Potential Link between Nuclear Winds and Cold Gas Outflows on Kiloparsec Scales in Reionization-Era Quasars}

\author[0000-0003-3307-7525]{Yongda Zhu}
\affiliation{Steward Observatory, University of Arizona, 933 North Cherry Avenue, Tucson, AZ 85721, USA}
\email[show]{yongdaz@arizona.edu}


\author[0000-0002-7893-6170]{Marcia J.\ Rieke}
\affiliation{Steward Observatory, University of Arizona, 933 North Cherry Avenue, Tucson, AZ 85721, USA}
\email{}

\author[0000-0001-6947-5846]{Luis C. Ho}
\affiliation{Kavli Institute for Astronomy and Astrophysics, Peking University, Beijing 100871, China}
\affiliation{Department of Astronomy, School of Physics, Peking University, Beijing 100871, China}
\email{}

\author[0000-0001-6561-9443]{Yang Sun}
\affiliation{Steward Observatory, University of Arizona, 933 North Cherry Avenue, Tucson, AZ 85721, USA}
\email{}

\author[0000-0003-2303-6519]{George H.\ Rieke}
\affiliation{Steward Observatory, University of Arizona, 933 North Cherry Avenue, Tucson, AZ 85721, USA}
\email{}

\author[0000-0003-3564-6437]{Feng Yuan}
\affiliation{Center for Astronomy and Astrophysics and Department of Physics, Fudan University, Shanghai 200438, China}
\email{}


\author[0000-0002-5268-2221]{Tom J. L. C. Bakx} 
\affiliation{Department of Space, Earth, \& Environment, Chalmers University of Technology, Chalmersplatsen 4 412 96 Gothenburg, Sweden}
\email{}

\author[0000-0003-2344-263X]{George D. Becker}
\affiliation{Department of Physics \& Astronomy, University of California, Riverside, CA 92521, USA}
\email{}

\author[0000-0001-5287-4242]{Jinyi Yang}
\affiliation{Department of Astronomy, University of Michigan, 1085 S. University, Ann Arbor, MI 48109, USA}
\email{}


\author[0000-0002-2931-7824]{Eduardo Ba\~nados}
\affiliation{Max-Planck-Institut f\"{u}r Astronomie, K\"{o}nigstuhl 17, D-69117 Heidelberg, Germany}
\email{}

\author[0000-0002-4314-021X]{Manuela Bischetti}
\affiliation{Dipartimento di Fisica ``Enrico Fermi'', Università di Pisa, Largo Bruno Pontecorvo 3, Pisa, I-56127, Italy}
\affiliation{INAF - Osservatorio Astronomico di Trieste, Via G. B. Tiepolo 11, I–34131 Trieste, Italy}
\email{}


\author[0000-0001-9420-7384]{Christopher Cain}
\affiliation{School of Earth and Space Exploration, Arizona State University, Tempe, AZ 85287-6004, USA}
\email{}

\author[0000-0003-3310-0131]{Xiaohui Fan}
\affiliation{Steward Observatory, University of Arizona, 933 North Cherry Avenue, Tucson, AZ 85721, USA}
\email{}

\author[0000-0001-7440-8832]{Yoshinobu Fudamoto} 
\affiliation{Center for Frontier Science, Chiba University, 1-33 Yayoi-cho, Inage-ku, Chiba 263-8522, Japan}
\affiliation{Steward Observatory, University of Arizona, 933 North Cherry Avenue, Tucson, AZ 85721, USA}
\email{}

\author[0009-0008-7862-5277]{Seyedazim Hashemi}
\affiliation{Department of Physics \& Astronomy, University of California, Riverside, CA 92521, USA}
\email{}

\author[0000-0002-2634-9169]{Ryota Ikeda}
\affiliation{Department of Astronomy, School of Science, SOKENDAI (The Graduate University for Advanced Studies), 2-21-1 Osawa, Mitaka, Tokyo 181-8588, Japan}
\affiliation{National Astronomical Observatory of Japan, 2-21-1 Osawa, Mitaka, Tokyo 181-8588, Japan}
\email{}

\author[0000-0001-7673-2257]{Zhiyuan Ji}
\affiliation{Steward Observatory, University of Arizona, 933 North Cherry Avenue, Tucson, AZ 85721, USA}
\email{}

\author[0000-0002-5768-738X]{Xiangyu Jin}
\affiliation{Steward Observatory, University of Arizona, 933 North Cherry Avenue, Tucson, AZ 85721, USA}
\email{}

\author[0000-0003-3762-7344]{Weizhe Liu}
\affiliation{Steward Observatory, University of Arizona, 933 North Cherry Avenue, Tucson, AZ 85721, USA}
\email{}

\author[0000-0003-4247-0169]{Yichen Liu}
\affiliation{Steward Observatory, University of Arizona, 933 North Cherry Avenue, Tucson, AZ 85721, USA}
\email{}

\author[0000-0002-6221-1829]{Jianwei Lyu}
\affiliation{Steward Observatory, University of Arizona, 933 North Cherry Avenue, Tucson, AZ 85721, USA}
\email{}

\author[0000-0002-5237-9433]{Hai-Xia Ma}
\affiliation{Division of Particle and Astrophysical Science, Nagoya University, Furo-cho, Chikusa-ku, Nagoya 464-8602, Japan}
\email{}

\author[0000-0001-8416-7673]{Tsutomu T.\ Takeuchi}
\affiliation{Division of Particle and Astrophysical Science, Nagoya University, Furo-cho, Chikusa-ku, Nagoya 464-8602, Japan}
\affiliation{The Research Center for Statistical Machine Learning, the Institute of Statistical Mathematics, 10-3 Midori-cho, Tachikawa, Tokyo 190-8562, Japan}
\email{}

\author[0000-0003-1937-0573]{Hideki Umehata}
\affiliation{Institute for Advanced Research, Nagoya University, Furocho, Chikusa, Nagoya 464-8602, Japan}
\affiliation{Department of Physics, Graduate School of Science, Nagoya University, Furocho, Chikusa, Nagoya 464-8602, Japan}
\email{}

\author[0000-0002-7633-431X]{Feige Wang}
\affiliation{Department of Astronomy, University of Michigan, 1085 S. University, Ann Arbor, MI 48109, USA}
\email{}

\author[0000-0003-0747-1780]{Wei Leong Tee}
\affiliation{Steward Observatory, University of Arizona, 933 North Cherry Avenue, Tucson, AZ 85721, USA}
\email{}

\begin{abstract}
\added{Feedback from accreting supermassive black holes may regulate galaxy evolution, but statistical evidence linking nuclear winds to kiloparsec-scale cold gas outflows remains limited in the early universe.}
Here we report statistical evidence for such a connection in a sample of luminous quasars at $z \sim 5.5$. We compare stacked [C\,{\sc ii}] 158 $\mu$m emission profiles from ALMA observations, which trace galactic-scale neutral gas, for quasars with and without broad absorption lines (BALs) that indicate powerful nuclear winds on sub-kiloparsec scales. \added{A total of 5 BAL and 11 non-BAL quasar spectra are included in the stacking analysis.} The BAL quasar stack exhibits a \added{potential} blueshifted broad component in the [C\,{\sc ii}] line profile, with a velocity offset of $\Delta v_{\rm b} = -2.1 \times 10^2\,\rm km\,s^{-1}$ and a full width at half maximum of $1.18 \times 10^3\,\rm km\,s^{-1}$, whereas the non-BAL stack shows \added{no obvious broad component}. \added{Using a conservative ``clean-stack'' selection that excludes quasars with partial [C\,{\sc ii}] spectral coverage, the BAL broad residual is reduced to a hint-level feature.} We estimate that \added{up to} a few percent to one-quarter of the nuclear wind energy may be transferred to cold neutral gas on kiloparsec scales. 
\added{Although the sample size is limited, these results suggest a potential link between BAL winds and cold gas feedback in quasar host galaxies.}
\added{These results provide empirical motivation for future tests of how multiphase outflows relate to the diversity of quasar host properties, including $M_{\rm BH}/M_*$.}
\end{abstract}

~

\section{Introduction}

Quasars are powered by accretion onto supermassive black holes (SMBHs). Understanding whether and how quasars impact their host galaxies is crucial for investigating the co-evolution of SMBHs and galaxies \citep{kormendy_coevolution_2013, sun_no_2025}, particularly in the early universe. In this context, it is essential to study how physics in the nuclear region influences the interstellar medium (ISM) on galactic scales.

The [C\,{\sc ii}] 158 $\mu$m fine-structure line is one of the strongest cooling lines of the ISM and serves as a key tracer of both star formation and gas dynamics \citep{stacey_158_1991, carilli_cool_2013, lagache_cii_2018}. It provides insights into the properties of the cold ISM, typically at temperatures of $T \sim 10^2$--$10^3$ K, and has been extensively used in studies of galaxies across cosmic time \citep{gullberg_nature_2015, jones_dynamical_2017}. Broad wings in the [C\,{\sc ii}] line profile have been interpreted as signatures of large-scale outflows driven by AGN or intense star formation \citep{cicone_very_2015, maiolino_evidence_2012}. However, recent re-analyses using deeper and wider-bandwidth data have questioned the statistical significance of such features in some individual systems \citep{meyer_physical_2022}. Moreover, star formation alone can also drive detectable [C\,{\sc ii}] wings, even in the absence of AGN activity, as shown in both stacked and individual high-redshift galaxies \citep{ginolfi_molecular_2017, ginolfi_alpine-alma_2020, herrera-camus_kiloparsec_2021, spilker_ubiquitous_2020, birkin_alma-cristal_2025}. These findings underscore the complexity of interpreting broad [C\,{\sc ii}] profiles and highlight the importance of carefully distinguishing between AGN- and star formation-driven feedback in population-level analyses.

Recent studies have explored whether [C\,{\sc ii}] outflows correlate with quasar properties such as luminosity or black hole accretion rate. Stacking analyses of $z > 6$ quasars have so far found limited evidence for outflow features in [C\,{\sc ii}] emission \citep{decarli_alma_2018, novak_no_2020, sawamura_no_2025}, while other works report more extended or kinematically disturbed [C\,{\sc ii}] components in selected individual quasars \citep{bischetti_multiphase_2024, carniani_alma_2018, izumi_subaru_2021}. These conflicting results suggest that detecting outflows using [C\,{\sc ii}] alone is nontrivial and may depend sensitively on sample properties, spatial resolution, and stacking methodology.

An effective tracer of nuclear winds is the presence of broad absorption line (BAL) features in quasar spectra. BALs are associated with high-velocity outflows launched near the accretion disk and are observed in a significant fraction of luminous quasars \citep{weymann_comparisons_1991, murray_accretion_1995, proga_dynamics_2000}. These outflows can reach velocities of thousands of kilometers per second and are thought to originate from radiation-driven winds in the nuclear region \citep{elvis_structure_2000}. \citet{bischetti_suppression_2022} reported that the fraction of BAL quasars increases toward higher redshift (e.g., in the XQR-30 program; \citep{dodorico_xqr-30_2023}), implying that nuclear winds may be more prevalent or visible during the reionization epoch. Investigating whether such small-scale winds are physically connected to larger-scale gas dynamics is crucial for understanding early SMBH--galaxy co-evolution. This is particularly relevant given recent evidence that some luminous quasars deviate from the local $M_{\rm BH}$--$M_*$ relation \citep{sun_m_-m_rm_2025}.

In this work, we investigate this connection using a sample of 17 quasars (16 used for stacking) at $z \sim 5.5$, near the end of reionization \citep{bosman_hydrogen_2022, zhu_long_2022}. Our observations, conducted with ALMA and Keck/ESI, enable a consistent comparison between BAL and non-BAL quasars. 
By stacking the [C\,{\sc ii}] profiles, we find that the BAL sample exhibits a statistically significant broad wing component not seen in the non-BAL stack, \added{while the conservative clean-stack analysis reduces the BAL broad residual to hint-level significance.} While this result suggests a potential coupling between nuclear and large-scale cold gas outflows, we interpret it in the context of existing uncertainties and complementary scenarios, including star formation-driven winds and orientation effects.

This paper is organized as follows. In Section~\ref{sec:data}, we describe the ALMA [C\,{\sc ii}] observations and Keck/ESI spectra of our quasar sample. Section~\ref{sec:results} presents the main results of our [C\,{\sc ii}] profile analysis. In Section~\ref{sec:discussion}, we discuss the implications of our findings, and we summarize our conclusions in Section~\ref{sec:summary}. Throughout this paper, we adopt the {\tt Planck18} cosmology \citep{planck_collaboration_planck_2020}, implemented in {\tt astropy} \citep{astropy_collaboration_astropy_2018}. All distances are quoted in proper units unless otherwise noted.

\begin{figure*}[!ht]
\centering
    \begin{overpic}[width=0.43\textwidth, keepaspectratio]{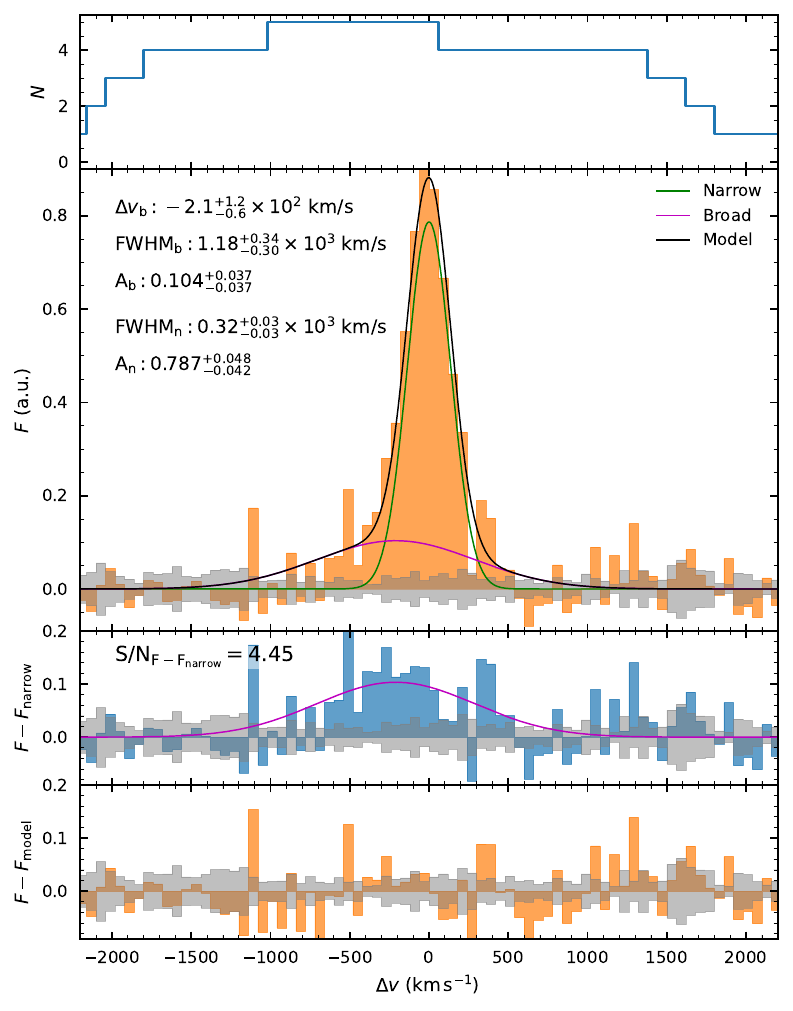}
        \put(15, 87){\textbf{(a) BAL stack}}
    \end{overpic}
    \begin{overpic}[width=0.43\textwidth, keepaspectratio]{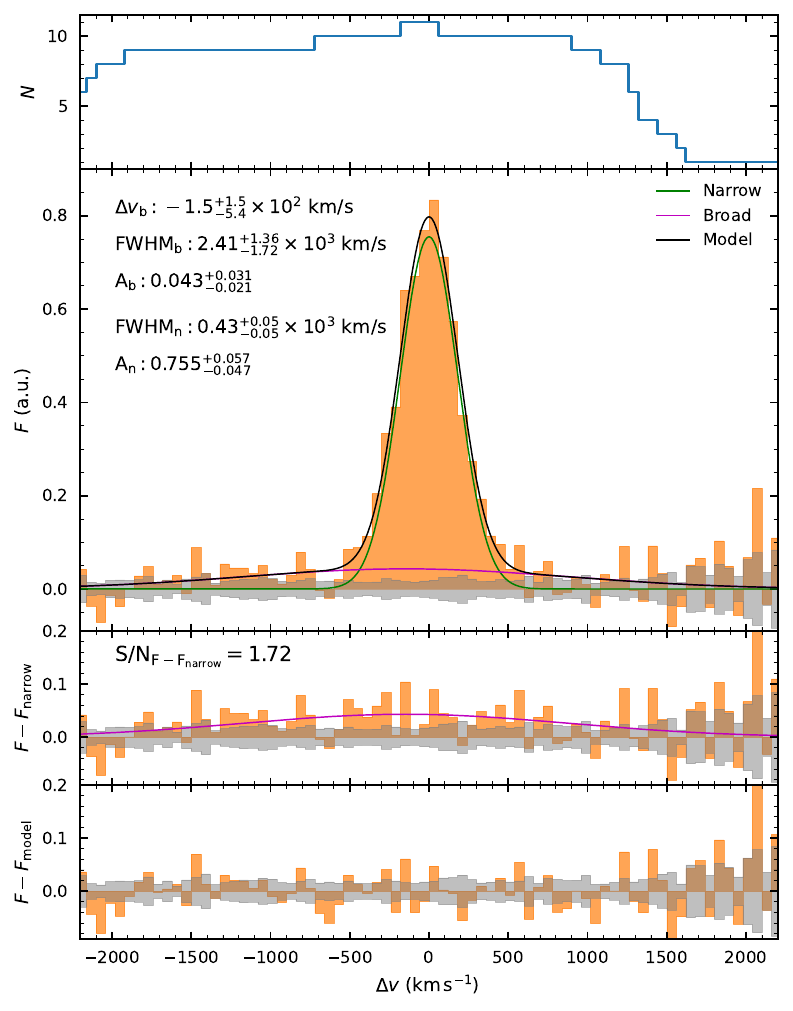}
        \put(15, 87){\textbf{(b) non-BAL stack}}
    \end{overpic}
    \begin{overpic}[width=0.43\textwidth, keepaspectratio]{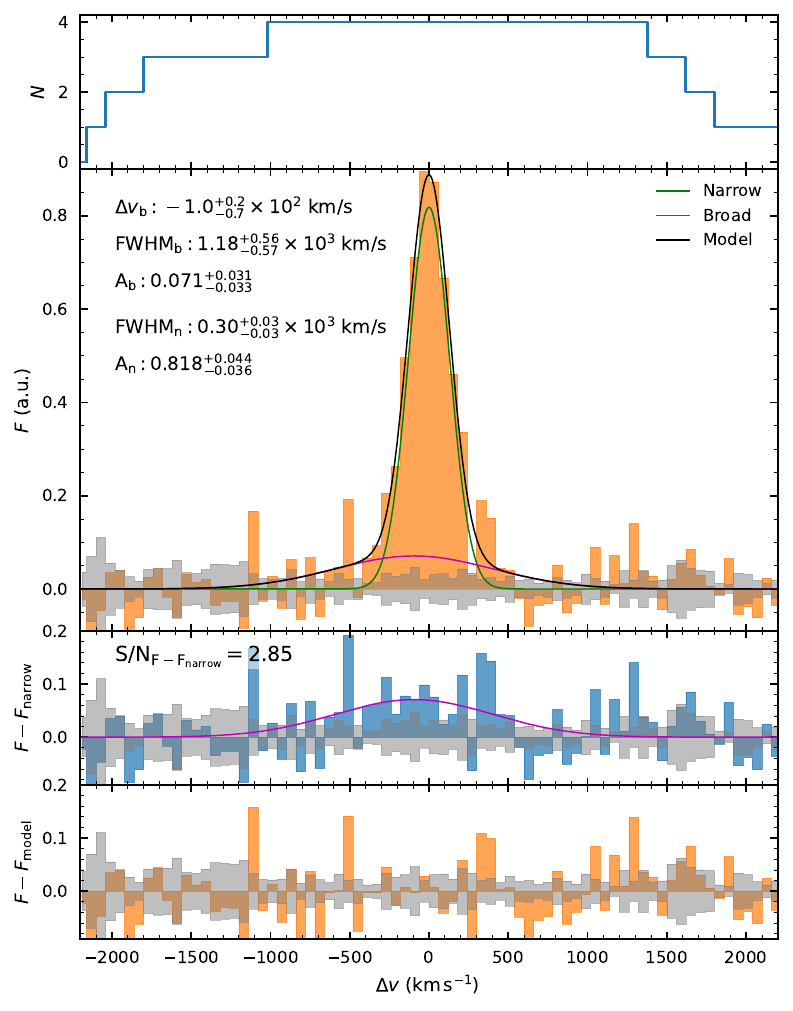}
        \put(15, 87){\textbf{(c) BAL stack -- clean}}
    \end{overpic}
    \begin{overpic}[width=0.43\textwidth, keepaspectratio]{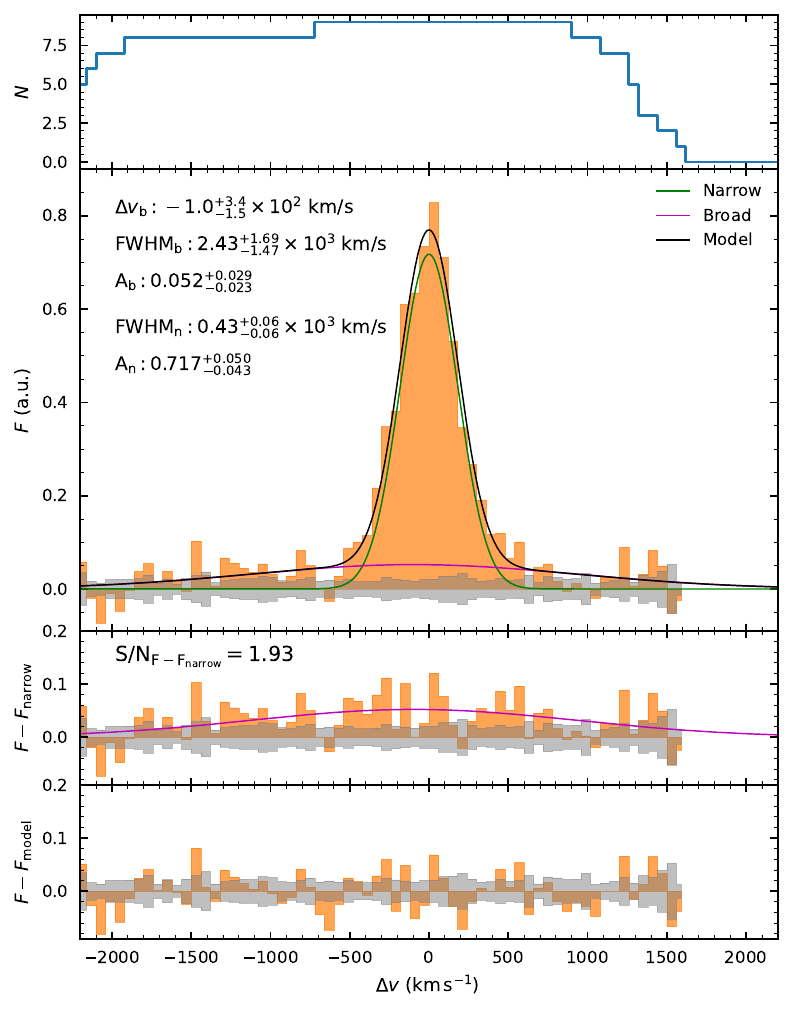}
        \put(15, 87){\textbf{(d) non-BAL stack -- clean}}
    \end{overpic}
    \caption{Stacked [C\,{\sc ii}] spectra.
    \textbf{(a)} Stacked [C\,{\sc ii}] profile for the BAL quasar sample. The top panel shows the number of spectra contributing to the stack at each velocity bin ($\Delta v$). The middle panel displays the normalized [C\,{\sc ii}] flux, with the best-fit narrow (green) and broad (magenta) Gaussian components. The combined model is shown in black, with a broad component centered at $\Delta v_{\rm b} = -2.1\times10^2 \, \rm km\,s^{-1}$ and a FWHM of $1.18\times10^3 \, \rm km\,s^{-1}$. The residuals between the total flux and the narrow component are shown in the third panel, highlighting the significance of the broad residual (mean S/N = 4.45 within the FWHM). The bottom panel shows the overall residuals after subtracting the total model.
    \textbf{(b)} Stacked [C\,{\sc ii}] profile for the non-BAL quasar sample. The profile shows no obvious broad residual, with residuals below 2$\sigma$.
    \added{Panels (c) and (d) provide a conservative ``clean-stack'' check excluding quasars with partial [C\,{\sc ii}] spectral coverage.}
    }
    \label{fig:stacked}
\end{figure*}

\section{Data}\label{sec:data}

We use the brightest quasars in the $z$-band ($m_{z} \lesssim 20$) selected by Yang et al.\ \citet{yang_discovery_2017, yang_filling_2019}, identified through a systematic search for high-redshift quasars. These quasars fill the so-called ``redshift gap'' at $z \sim 5.5$, a regime that was historically underrepresented in quasar surveys due to color degeneracy with brown dwarfs in broadband photometry. Yang et al.\ overcame this challenge by incorporating additional $Y$-band imaging to break the degeneracy, enabling us to investigate quasar properties in this key transitional epoch near the end of cosmic reionization.

~

\subsection{ALMA Observations}

In Cycle 9 (2022.1.00662.S; PI: Zhu), we conducted ALMA Band 7 observations targeting 21 quasars at $z \sim 5.5$ from \citet{yang_discovery_2017,yang_filling_2019}, aiming to determine precise systemic redshifts ($\Delta z \sim 10^{-4}$) through [C\,{\sc ii}] 158$\mu$m emission (details in \cite{zhu_probing_2023}). Each observation employed two overlapping spectral windows fully covering the [C\,{\sc ii}] line at the expected redshift, alongside two additional spectral windows for dust continuum measurements. The observations used the C43-(1, 2, and 3) configurations, achieving an angular resolution of $\sim 1''$ across the entire sample.

Data calibration and reduction were performed using the CASA pipeline (version 6.4.1.12; \cite{mcmullin_casa_2007, casa_team_casa_2022}). We generated data cubes and imaged the [C\,{\sc ii}] emission following procedures outlined in \citet{eilers_detecting_2020}. Continuum subtraction was conducted in the $uv$-domain using all line-free channels across the four spectral windows by fitting a first-order polynomial, \added{and all stacked spectra in this work are therefore based on continuum-subtracted data}. Imaging was performed with {\tt tclean}, using natural weighting to optimize sensitivity. The average RMS noise level across the dataset is $\sim 0.25\ {\rm mJy\,beam^{-1}}$ per 30 MHz bin.

Spectra were extracted using a single-beam aperture centered on each source to minimize contamination from extended [C\,{\sc ii}] emission and complex host galaxy kinematics. To evaluate the impact of aperture size, we also tested extractions using a standard 2-beam diameter aperture. For the BAL quasar stack, the broad wing detection significance decreased from 4.45 (Section~\ref{sec:results}) to 3.3$\sigma$, while for the non-BAL stack, it dropped from $<$2$\sigma$ to below 1.5$\sigma$. Although a larger aperture introduces increased noise, the decrease in S/N for the BAL stack exceeds what is expected from noise alone, suggesting that physical dilution (via inclusion of unrelated kinematic components) is the dominant factor. This supports our choice of using the single-beam aperture for the fiducial analysis.

We include only quasars with successful [C\,{\sc ii}] detections. Three quasars --- J0157+3001, J1420-1602, and J1527+0641 --- are excluded because the [C\,{\sc ii}] line may fall outside the spectral coverage. Additionally, J1133+1603 is excluded due to its close companion exhibiting a [C\,{\sc ii}] bridge suggesting interaction and complex gas kinematics \citep{zhu_discovery_2024}. The final ALMA sample comprises 17 quasars. The resulting [C\,{\sc ii}] images and spectra are presented in Figure \ref{fig:cii}. We determined the systemic redshift for each quasar by fitting a Gaussian profile to the [C\,{\sc ii}] emission line.

~

\subsection{ESI Spectra}

We acquired the Keck/ESI spectra for most quasars in this sample, which we use to identify BAL features in the rest frame UV. We conducted ESI observations for 13 quasars between 2021 and 2024 and retrieved archival spectra for one additional quasar. For the remaining three (J1006, J1048, and J1614), we used existing confirmation spectra from \citet{yang_discovery_2017,yang_filling_2019}, obtained with SSO-2.3m/WiFeS \citep{dopita_wide_2007,dopita_wide_2010} and Palomar-200-inch (P200)/DBSP.

The data were reduced using a custom IDL/GDL\citep{coulais_status_2010} pipeline, following procedures described in \citet{becker_evolution_2019} and \citet{zhu_probing_2023}. This pipeline incorporates optimal sky subtraction techniques \citep{kelson_optimal_2003}, one-dimensional spectral extraction \citep{horne_optimal_1986}, and telluric absorption corrections based on the Cerro Paranal Advanced Sky Model \citep{noll_atmospheric_2012, jones_advanced_2013}. The extracted spectra have a pixel size of 15 $\rm km\, s^{-1}$, with a typical velocity resolution of approximately 45 $\rm km\, s^{-1}$ (FWHM). The signal-to-noise ratio (S/N) per 30 $\rm km\, s^{-1}$ bin is greater than 10 near 1285 \AA\ in the rest frame.

In this work, we focus on quasars with BAL outflows, i.e., classic BAL quasars, defined by a balnicity index ${\rm BI}_0 > 0\ \rm km\, s^{-1}$ \citep{weymann_comparisons_1991, gibson_catalog_2009}. The balnicity index is computed using the equation:

\begin{equation}
    {\rm BI}_0 = \int_{v_{\rm min}}^{v_{\rm max}} \left( 1 - \frac{f(v)}{0.9} \right) C(v) dv,
\end{equation}
\noindent
where $f(v)$ is the normalized flux as a function of velocity, $C(v)$ is a continuity function equal to 1 when absorption depth exceeds 10\% over at least 2000 km s$^{-1}$, and 0 otherwise. We adopt $v_{\rm min} = 0$ km s$^{-1}$ and $v_{\rm max} = 64500$ km s$^{-1}$ \citep{bischetti_suppression_2022}.

We set zero velocity at C\,{\sc IV} and identify BALs in both Si\,{\sc IV} and C\,{\sc IV}. Due to incomplete wavelength coverage, we do not attempt precise ${\rm BI}_0$ measurements. A major source of uncertainty in computing ${\rm BI}_0$ is the power-law continuum fitting. Three quasars (J1006, J1048, J1650) yield ${\rm BI}_0 > 0$, but their BAL features are not visually apparent and may be influenced by telluric absorption or noise near spectral edges. To ensure robustness, we classify only quasars with significant BAL features ($\rm BI_0 > 1000\ \rm km\, s^{-1}$) as BAL quasars. We identify five quasars meeting this criterion. Among our sample, J0012, J2317, and J2325 exhibit the most unambiguous BAL features ($\rm BI_0 > 2500\ \rm km\, s^{-1}$). As a consistency check, stacking [C\,{\sc ii}] emission for these three quasars alone yields a significant ($>3\sigma$) broad [C\,{\sc ii}] wing (Section \ref{sec:results}).

To compute $f(v)$, we normalized observed fluxes by fitting a power-law continuum to each quasar spectrum in the rest-frame wavelength range of 1285--1450 \AA, masking strong emission or absorption lines. Figure \ref{fig:ESI} displays all ESI spectra, continuum fits, and BAL identifications. The detailed properties of our quasar sample are summarized in Table \ref{tab:QSOlist}.

\section{Stacked [C II] Spectra}\label{sec:results}

We create stacked [C\,{\sc ii}] spectra for the five strong BAL quasars and eleven non-BAL quasars, respectively. The quasar J0056+2241 is excluded from both stacks due to its independently identified strong and extended [C\,{\sc ii}] feature in the emission line map. To construct the stacks, we normalize each spectrum by its peak flux, shift the [C\,{\sc ii}] emission line to the systemic redshift, and compute the mean stack in velocity space ($\Delta v$ --- the velocity offset relative to the systemic redshift) using bins of 60 $\rm km\,s^{-1}$. The spectra are weighted by their inverse variance, although we verified that not weighting by inverse variance does not change the results, given the relatively consistent rms level across the entire quasar sample. We perform 1,000 bootstrap resamplings (with replacement) of 5 spectra from the BAL sample and 11 spectra from the non-BAL sample, respectively, to estimate the variance in the stacked profiles. \added{Throughout this paper, we classify a stacked feature as a significant detection if its mean S/N is $>3\sigma$, as a hint if $2\sigma<{\rm S/N}<3\sigma$, and as no obvious feature if ${\rm S/N}<2\sigma$.}

For the stacked spectra, we find that a single Gaussian profile centered at $\Delta v=0$ cannot adequately describe the signal, especially for the BAL stack. Therefore, we fit a two-component Gaussian model: a narrow component with velocity dispersion $\sigma < 500 \, \rm km\,s^{-1}$ and a broad component with $\sigma > 500 \, \rm km\,s^{-1}$. The narrow component's line center is fixed at $\Delta v = 0$, while the broad component's center is allowed to vary. We derive uncertainties in the fit parameters based on the stacked fluxes across the bootstrap realizations.

Figure \ref{fig:stacked} panels (a) and (b) present the stacked [C\,{\sc ii}] profiles for the BAL and non-BAL samples, respectively. In the BAL stack, a single narrow Gaussian cannot fully explain the observed profile due to the presence of a blueshifted broad component. The residual broad emission is detected with a mean signal-to-noise ratio of 4.45 within its FWHM (4.45$\sigma$
\footnote{However, the $\sigma$ value quoted here represents the signal detection level rather than a strict statistical significance, as there remains a $\lesssim 2\%$ probability that a stack of five non-BAL quasars could yield a similar broad component detection (see Section \ref{sec:discussion} and Figure \ref{fig:small-sample}).}
). The best-fit Gaussian for the broad component yields a line center at $\Delta v_{\rm b} = -2.1_{-0.7}^{+1.3}\times 10^2 \, \rm km\,s^{-1}$ and a FWHM of $1.18_{-0.35}^{+0.39} \times 10^3 \, \rm km\,s^{-1}$. We revisit the individual spectra in Figure \ref{fig:cii} and notice that some BAL quasars show a slight gradual decrease in the flux toward the bottom of their single Gaussian fit; however, these features are too noisy to make robust measurements individually (e.g., J0012 and J2325). Nevertheless, these features become prominent in the stacked spectrum.

In contrast, the non-BAL stack does not exhibit a significant broad component, and the broad residual has a detection significance below $2\sigma$. The Gaussian fit for the broad component is unconstrained. We note that the noise level in the non-BAL stack is lower than in the BAL stack due to the larger sample size. 
\added{This implies that the non-BAL stack shows no obvious broad residual based on our S/N classification.}

Additionally, we note that the narrow component of the BAL stack is narrower ($\rm FWHM = 3.2\times10^2\,km\,s^{-1}$) than that of the non-BAL case ($\rm FWHM = 4.3\times10^2\,km\,s^{-1}$). This difference might be attributed to viewing angle effects (see Section~\ref{sec:discussion}).

\added{We further construct ``clean'' stacks by excluding quasars with only partial spectral coverage of the [C\,{\sc ii}] line (J0012 in the BAL sample; J1006 and J1016 in the non-BAL sample).}
\added{The resulting clean BAL and non-BAL stacks are shown in Figure~\ref{fig:stacked} panels (c) and (d), respectively. In the clean BAL stack, the broad residual is reduced to ${\rm S/N}=2.85$, corresponding to a hint-level feature under our S/N classification, while the clean non-BAL stack shows no obvious broad residual with ${\rm S/N}=1.93$. We use these clean-stack results as a conservative check on the robustness of the broad component and to guide the interpretation in the remainder of the paper.}

\section{Discussion}\label{sec:discussion}
\subsection{Nuclear Winds and the Origin of the [C\,{\sc ii}] Wing}

The observed association between [C\,{\sc ii}] outflows and BAL quasars may be \added{consistent with} coupling between parsec-scale nuclear winds and galactic-scale cold gas. Multi-phase simulations predict such coupling \citep{tanner_simulations_2022}, and our stacking analysis provides \added{suggestive, statistical evidence} for \added{a link} at high redshift. \added{Figure~\ref{fig:stacked}(c) shows the clean BAL stack, which exhibits a hint-level blueshifted broad residual.} \added{If the BAL outflow axis aligns with our line of sight, we observe blueshifted broad absorption features in high-ionization UV lines (e.g., C\,{\sc iv}, Si\,{\sc iv}) and may also preferentially detect blueshifted [C\,{\sc ii}] emission on kiloparsec scales tracing disturbed or outflowing neutral gas.} In contrast, non-BAL quasars, either lacking strong nuclear winds or viewed from different sightlines, may show narrower or no \added{significant} blueshifted [C\,{\sc ii}] wings. The narrower width of the [C\,{\sc ii}] narrow component in the BAL stack (FWHM = $3.2 \times 10^2$\,km\,s$^{-1}$) compared to the non-BAL stack (FWHM = $4.3 \times 10^2$\,km\,s$^{-1}$) is consistent with a more face-on orientation, while broader [C\,{\sc ii}] lines in non-BALs likely reflect complex rotational and dynamical structures viewed closer to edge-on.

Alternative interpretations must be considered. First, galaxy mergers or unresolved companion galaxies may broaden [C\,{\sc ii}] profiles or introduce asymmetric wings. Although our sample excludes obvious companions in the ALMA cubes, higher-resolution observations would be needed to fully rule out blended structures. Second, star formation-driven outflows are known to contribute to [C\,{\sc ii}] wings, even in non-AGN systems \citep{ginolfi_molecular_2017, ginolfi_alpine-alma_2020, herrera-camus_kiloparsec_2021, spilker_ubiquitous_2020, birkin_alma-cristal_2025}. Since quasars often reside in star-forming galaxies, some [C\,{\sc ii}] broad emission may arise from stellar feedback rather than AGN winds. However, \added{if star formation were the dominant cause, we would expect similarly significant [C\,{\sc ii}] wings in the non-BAL sample, which are not observed at comparable significance.} Still, we caution that our stacking analysis cannot unambiguously disentangle AGN- and star formation-driven outflows. The energetics discussed below therefore represent \added{illustrative upper limits on} AGN coupling efficiency, assuming the [C\,{\sc ii}] broad component is primarily AGN-driven.

Third, orientation effects likely play a significant role. BALs are preferentially observed along sightlines aligned with the nuclear wind axis \citep{elvis_structure_2000}, \added{as they are detected along the line of sight to the quasar nucleus}. \added{By contrast, the [C\,{\sc ii}] emission associated with large-scale outflows arises from potentially offset, spatially extended gas in the host galaxy and is not confined to the nuclear sightline.} Large-scale outflows, whether AGN- or star formation-driven, are expected to emerge perpendicular to the galactic plane. This implies a potential geometric alignment bias: face-on systems are more likely to show both BAL features and outflow signatures in integrated spectra. The narrower [C\,{\sc ii}] profiles in BAL quasars further support this orientation scenario. However, disk--AGN alignment is not guaranteed and may vary across systems \citep{stanley_spectral_2019}. The observed trend could also arise from structural alignment between the quasar outflow axis and the host disk normal. A more complete treatment of orientation and multiphase outflows will require spatially resolved data, ideally combining [C\,{\sc ii}], ionized gas, and stellar continuum mapping.

While our results may support a connection between BAL winds and kiloparsec-scale cold gas outflows, we stress that other mechanisms—including star formation feedback, minor mergers, and viewing angle—likely contribute. A definitive causal link between nuclear winds and [C\,{\sc ii}] broad wings will require larger samples and spatially resolved observations to isolate these effects.

~

\subsection{Energetics and Coupling Efficiency Between BAL and Cold Gas Outflows}

To quantify the potential coupling between nuclear and cold gas outflows, we estimate the kinetic power of each phase. For the BAL outflows, we adopt a hydrogen column density of $N_{\rm H} = 10^{20.1\sim 22.6}\ \rm cm^{-2}$ \citep{xu_vltx-shooter_2019} and outflow radii $\log_{10} (R_{\rm out}/{\rm pc}) = 0 \sim 2$ \citep{liu_galactic_2022, veilleux_galactic_2022, bischetti_multiphase_2024}. With a mean velocity offset of the broad absorption troughs of $v_{\rm BAL} = 2.5\times 10^4\,\rm km\,s^{-1}$ and assuming covering fraction of $f_{\rm cov} = 0.2$ \citep{hewett_frequency_2003}, the mass outflow rate is

\begin{equation}
    \dot{M}_{\rm BAL} = 4 \pi R_{\rm out} N_{\rm H} \mu m_{\rm p} f_{\rm cov} v_{\rm BAL},
\end{equation}
where $\mu = 1.4$ accounts for helium. The corresponding kinetic power is

\begin{equation}
    \dot{E}_{\rm BAL} = \frac{1}{2} \dot{M}_{\rm BAL} v_{\rm BAL}^2
    \approx 10^{44.3-46.3} \rm erg\,s^{-1}\,.
\end{equation} 

For the [C\,{\sc ii}] outflow, we adopt an outflow radius of 3.0 kpc, matching the mean ALMA beam size and consistent with the measurements in \citet{bischetti_widespread_2019}. The fraction of the flux in the broad component over the total stacked flux is $F_{\rm broad} / F_{\rm total} = 0.32 \pm 0.11$. The outflow velocity is computed in the standard way following e.g., \citet{bischetti_widespread_2019}:

\begin{equation}
    v_{\rm [CII]} = |\Delta v_{\rm b}| + \frac{\rm FWHM_{\rm broad}}{2} = 
    (8.0 \pm 2.1) \times 10^2\ \rm km\,s^{-1}.
\end{equation}

Following \citet{hailey-dunsheath_detection_2010}, the outflow mass is given by

\begin{equation}
    \begin{aligned}
    M_{\rm out} / M_\odot & = 0.77 \left( \frac{0.7L_{\rm [CII]}}{L_\odot} \right) \left( \frac{1.4 \times 10^{-4}}{X_{\rm C^+}} \right) \\
                           & \times \frac{1 + 2e^{-91K/T} + n_{\rm crit}/n}{2e^{-91K/T}},
    \end{aligned}
\end{equation}
where $L_{\rm [CII]}$ is the [C\,{\sc ii}] luminosity of the broad component determined by $F_{\rm broad} / F_{\rm total}$, $n_{\rm crit} \sim 3 \times 10^3\ \rm cm^{-3}$ is the critical density, and $X_{\rm C^+} = 10^{-4}$ is the assumed abundance of C$^+$ \citep{maiolino_evidence_2012, bischetti_widespread_2019}. Assuming $n \gg n_{\rm crit}$ and a temperature range of $T \sim 100$--$1000$ K, we obtain a mean outflow mass of $M_{\rm out} = (9.3 \pm 0.2) \times 10^8 M_\odot$.

The time-averaged mass outflow rate is then given by

\begin{equation}
    \dot{M}_{\rm out} = \frac{v_{\rm [CII]} \times M_{\rm out}}{R_{\rm out}}
    \approx 10^{2.3-2.4} M_\odot\,\rm yr^{-1}\,,
\end{equation} 
and the kinetic power by
\begin{equation}
    \dot{E}_{\rm out} = \frac{1}{2} \dot{M}_{\rm out} \times v_{\rm [CII]}^2 \approx 10^{43.6 - 43.8}\, \rm erg\,s^{-1}.
\end{equation} 


Finally, the coupling efficiency between the BAL and [C\,{\sc ii}] outflows is

\begin{equation}
    \eta = \frac{\dot{E}_{\rm out}}{\dot{E}_{\rm BAL}} = 0.025_{-0.022}^{+0.23} \, .
\end{equation} 

\added{Interpreting the stacked broad component as an outflow, the inferred coupling efficiency is of order $\eta\sim10^{-2}$, but should be regarded as an illustrative, order-of-magnitude estimate. The BAL-wind energetics depend on uncertain parameters (e.g., $N_{\rm H}$, $R_{\rm out}$, and $f_{\rm cov}$), and the relevant physical scales differ substantially between the nuclear BAL region and the kiloparsec-scale gas traced by [C\,{\sc ii}] \citep{fiore_agn_2017, veilleux_cool_2020}. In addition, the [C\,{\sc ii}] broad component traces only one phase of a potentially multiphase outflow, and may reflect time-averaged or past activity rather than being strictly contemporaneous with the current BAL phase.}
\added{Moreover, since the conservative clean-stack analysis yields a weaker broad residual, the energetics and $\eta$ inferred from the full-stack broad component should be regarded as approximate upper limits.}

\added{Under the same assumptions, the implied cold-gas mass outflow rate is $\dot{M}_{\rm out}\sim10^2\,M_\odot\,{\rm yr^{-1}}$, comparable to the star formation rates inferred for luminous quasar hosts at these redshifts. However, translating this into long-term quenching requires additional constraints on the host potential, gas recycling, and the quasar duty cycle.}

\added{The BAL fraction among luminous quasars at $z \gtrsim 5.5$ is high (approximately 50\%; \citep{bischetti_suppression_2022}). Some high-redshift quasars also exhibit elevated $M_{\rm BH}/M_*$ ratios relative to the local Magorrian relation, though these deviations can be explained by rapid black hole growth, delayed stellar mass assembly, or selection biases \citep{sun_m_-m_rm_2025}. The role of BAL-driven feedback remains uncertain, and recent observations suggest that star formation can be enhanced in some quasar hosts \citep{molina_enhanced_2023}.}
\added{Overall, our stacking results are consistent with a scenario in which nuclear winds and large-scale cold-gas kinematics are linked in at least a subset of luminous quasars, but the current evidence remains statistical and does not establish a causal connection. An additional caveat is that stacked spectra can be affected by companion galaxies or merging components within the ALMA beam, which could mimic a distinct velocity component in unresolved [C\,{\sc ii}] profiles. Future spatially resolved observations, together with larger and more uniform samples, will be essential for separating outflows from companions and for assessing whether any coupling between BAL winds and host-scale cold gas is generic.}


\begin{figure}[!ht]
    \centering
    \includegraphics[width=1\linewidth]{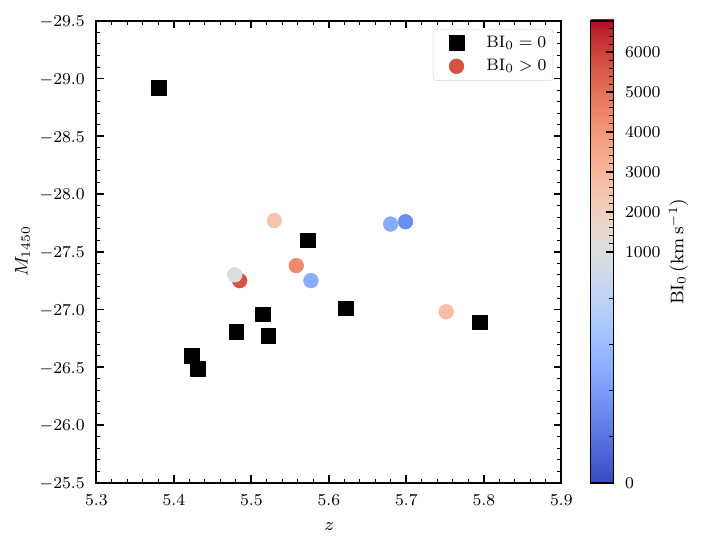}
    \caption{UV magnitude ($M_{1450}$) versus redshift for the quasars used in this work. The five quasars with strong BAL features (${\rm BI}_0 > 1000$ km\,s$^{-1}$) are used to create the stacked [C\,{\sc ii}] profile shown in Figure~\ref{fig:stacked}(a).}
    \label{fig:M1450}
\end{figure}

\begin{figure}[!ht]
    \centering
    \includegraphics[width=0.95\linewidth]{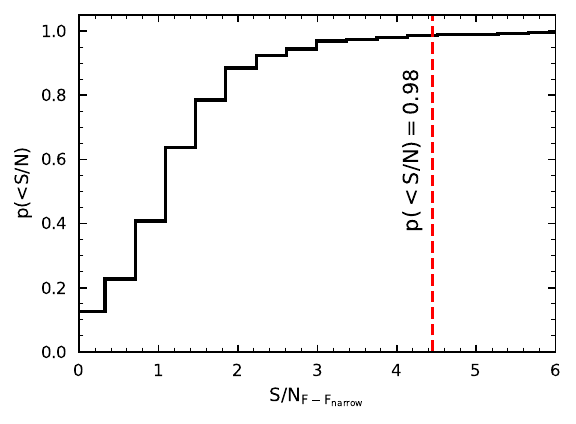}
    \caption{Cumulative distribution of the S/N of the broad component based on random stacks of five non-BAL quasars. 
    The probability of reproducing an S/N=4.4 detection of the broad [C\,{\sc ii}] component in non-BAL quasars is less than 0.02. \added{This null test disfavors pure-noise explanations for the full-stack BAL broad residual, although the conservative clean-stack analysis yields only a hint-level feature.} \added{Systematic effects related to data reduction or continuum subtraction are not an obvious explanation for the full-stack residual.}
    }
    \label{fig:small-sample}
\end{figure}

\begin{figure*}[!ht]
    \centering
    \includegraphics[width=1.0\linewidth]{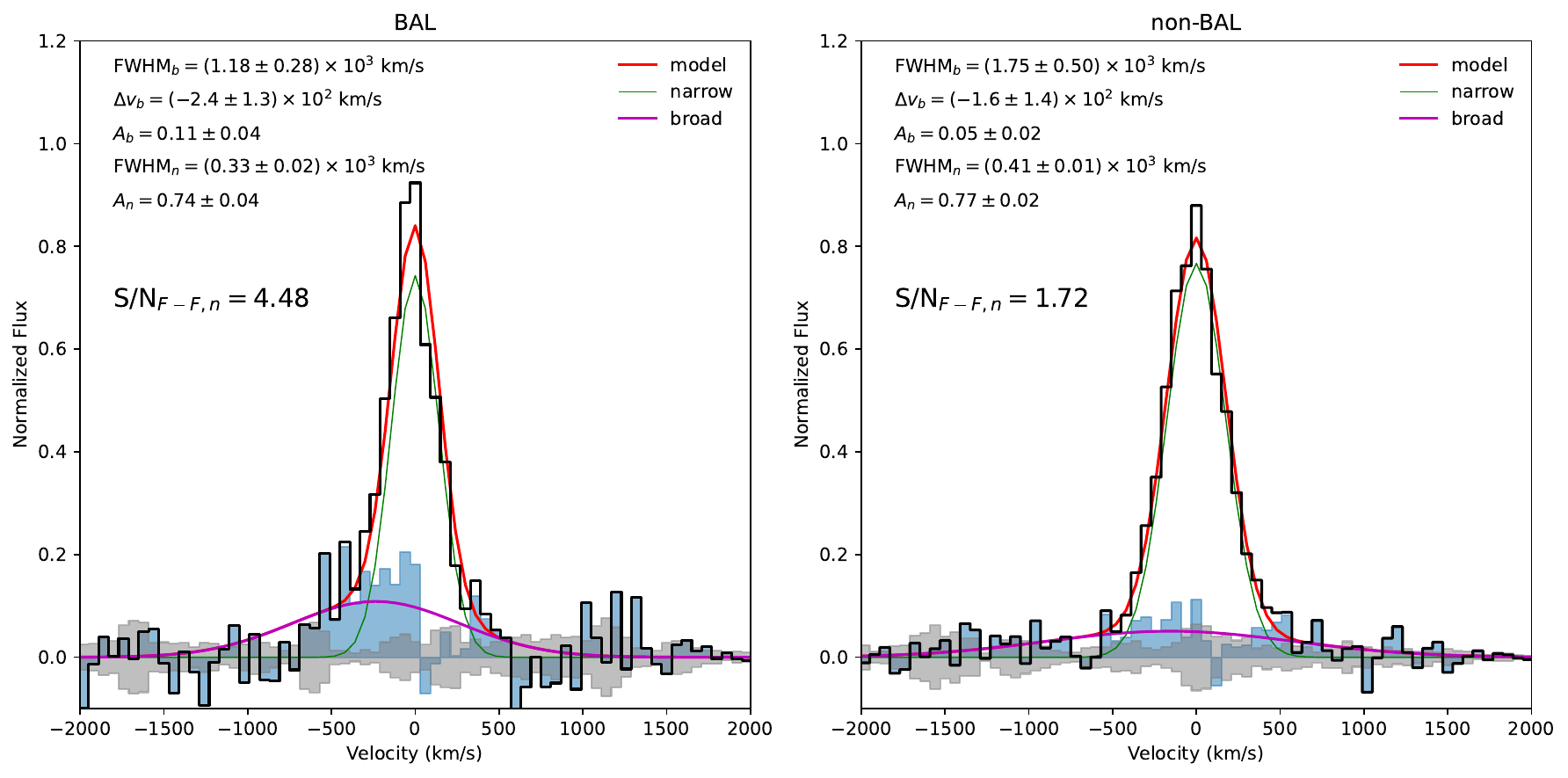}
    \caption{[C\,{\sc ii}] emission from the stacked image cubes of BAL (left panel) and non-BAL (right panel) quasars. The black curves represent the normalized flux (arbitrary units), while the blue shaded regions show the residual flux after subtracting the narrow component. The gray shaded regions indicate the $\pm 1\sigma$ noise level. The fitting parameter uncertainties shown in this plot are derived from a single stack, rather than the bootstrap approach used in Figure \ref{fig:stacked}. When stacking image cubes instead of 1D spectra, the BAL sample still exhibits a significant broad component at $4.48\sigma$, whereas the non-BAL sample remains below the $2\sigma$ detection threshold.}
    \label{fig:cube-stack}
\end{figure*}

\begin{figure*}[!ht]
\centering
\begin{minipage}[b]{0.31\textwidth}
  \includegraphics[width=\linewidth]{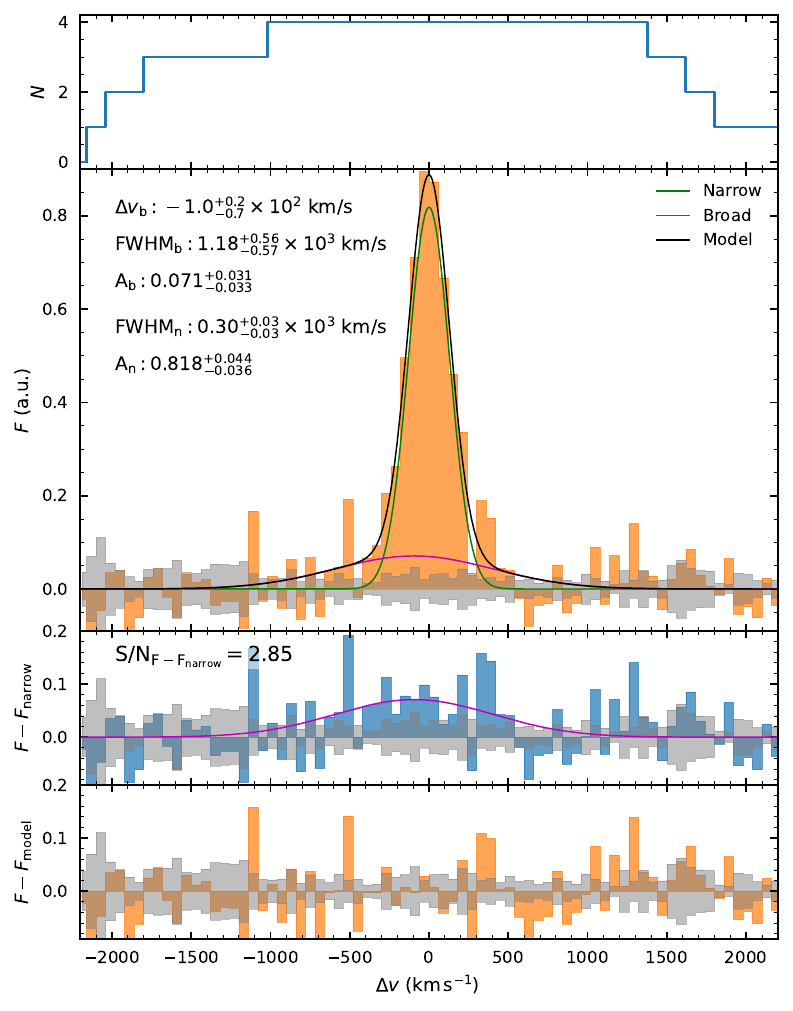}
  \centering (a) Excluding J0012
\end{minipage} \hfill
\begin{minipage}[b]{0.31\textwidth}
  \includegraphics[width=\linewidth]{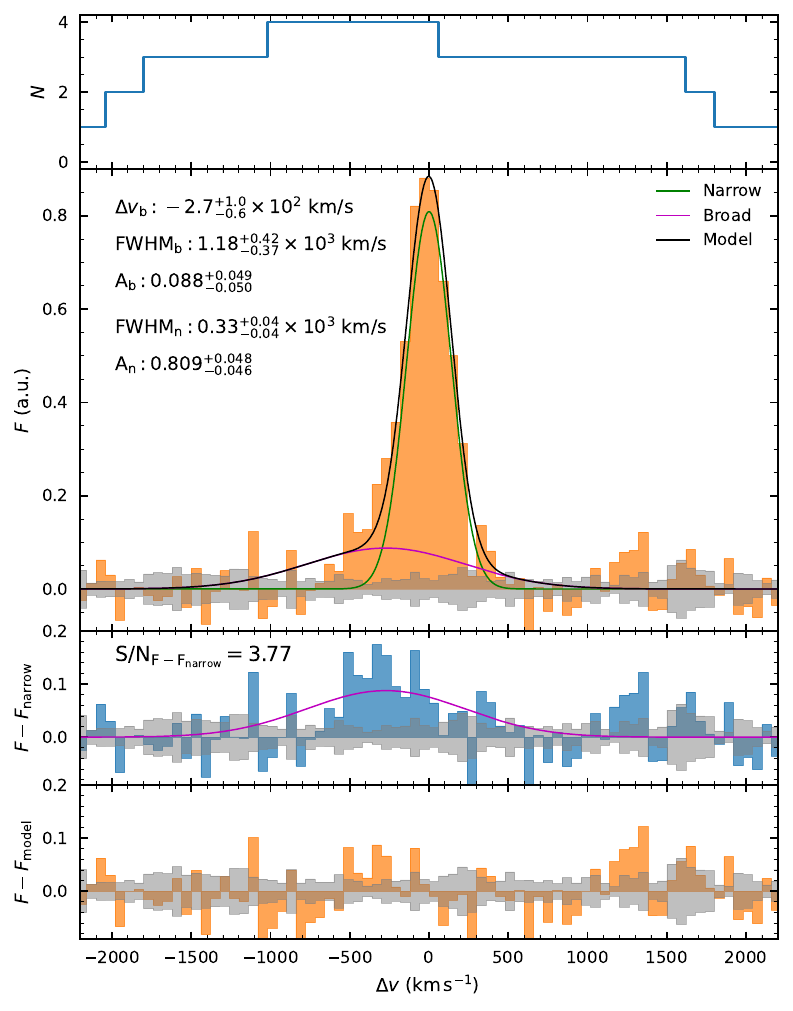}
  \centering (b) Excluding J1022
\end{minipage} \hfill
\begin{minipage}[b]{0.31\textwidth}
  \includegraphics[width=\linewidth]{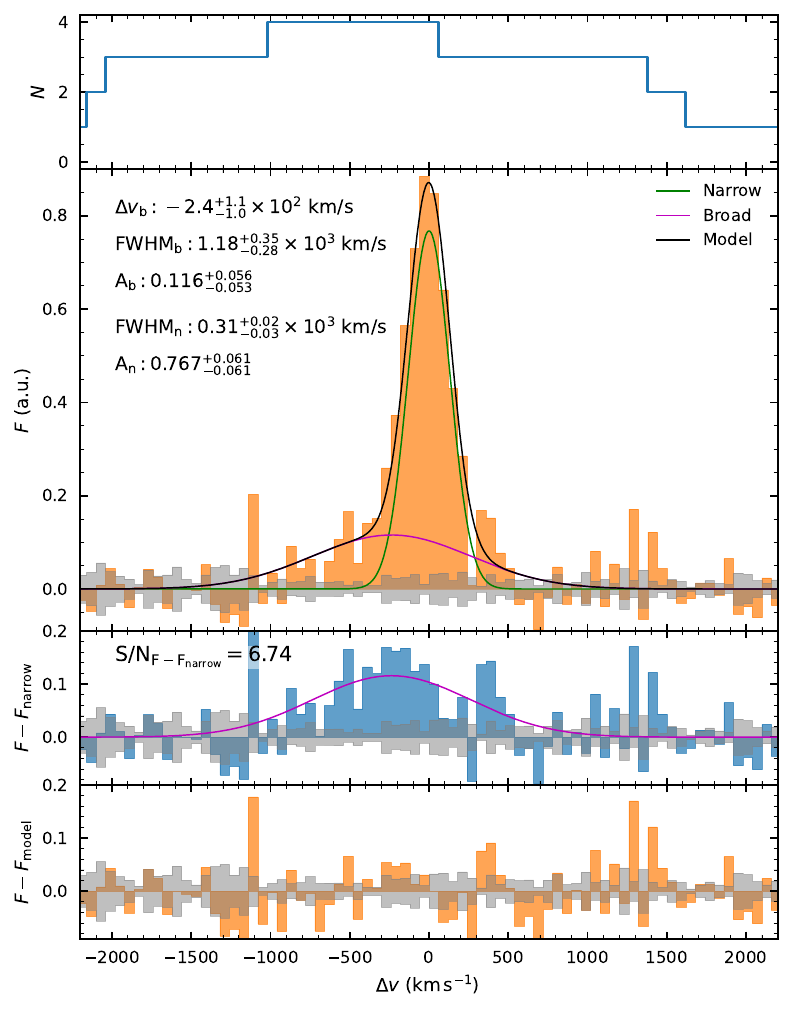}
  \centering (c) Excluding J2207
\end{minipage}
\vspace{0.4cm}
\begin{minipage}[b]{0.31\textwidth}
  \includegraphics[width=\linewidth]{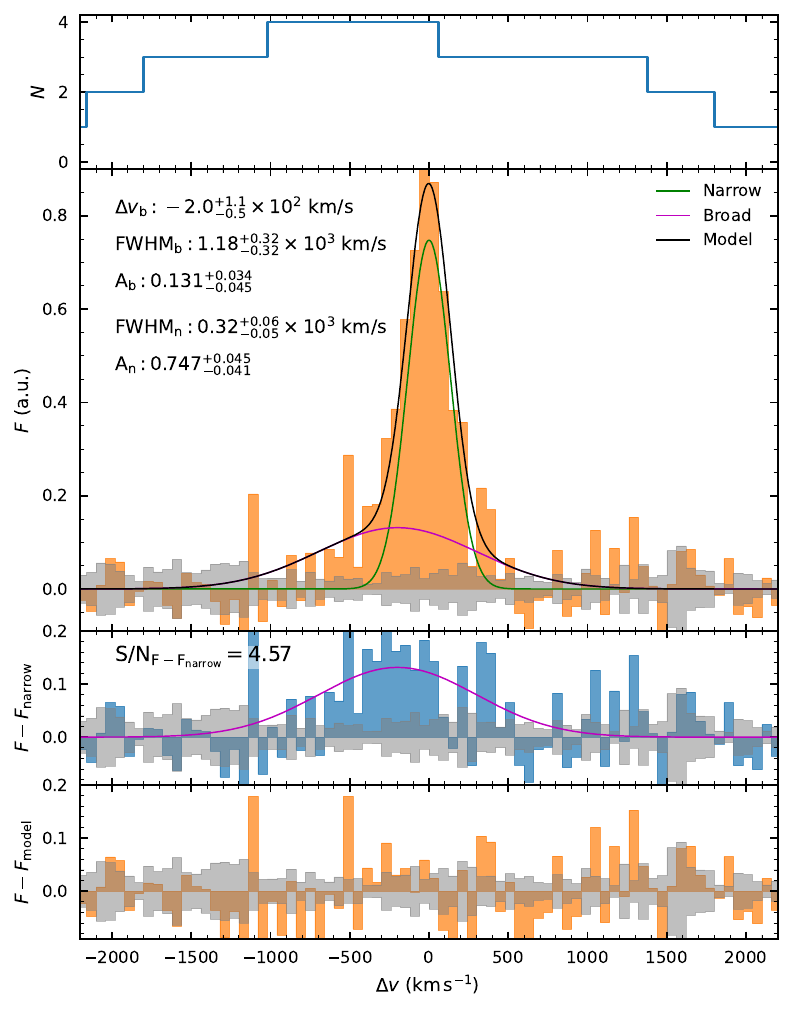}
  \centering (d) Excluding J2317
\end{minipage} \hfill
\begin{minipage}[b]{0.31\textwidth}
  \includegraphics[width=\linewidth]{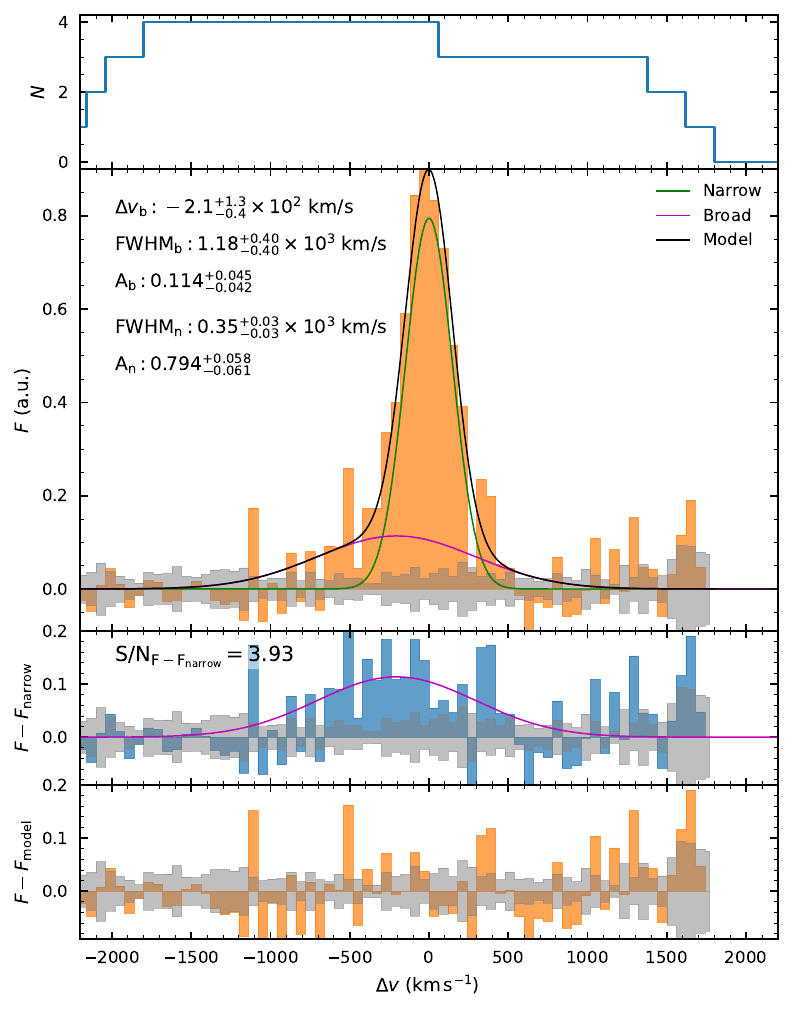}
  \centering (e) Excluding J2325
\end{minipage}
\caption{Jackknife tests of the [C\,{\sc ii}] stacked spectrum for the BAL QSO sample. Each panel shows the stack of 4 out of 5 BAL quasars, with one excluded. The broad component remains visible across all tests, \added{but its S/N decreases when excluding J0012,} \added{indicating that the result is not solely driven by a single source.}}
\label{fig:jackknife}
\end{figure*}

\begin{figure*}[!ht]
\centering
\begin{minipage}[b]{0.45\textwidth}
  \includegraphics[width=\linewidth]{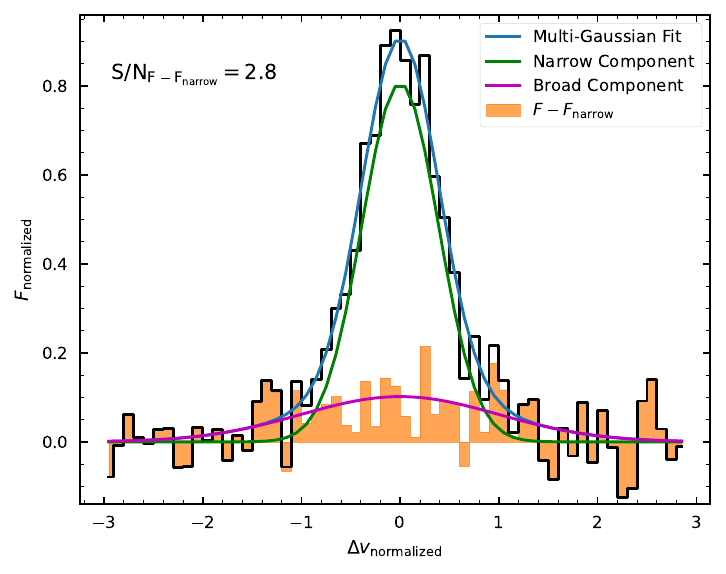}
  \centering (a) BAL stack (velocity-normalized)
\end{minipage} \hfill
\begin{minipage}[b]{0.45\textwidth}
  \includegraphics[width=\linewidth]{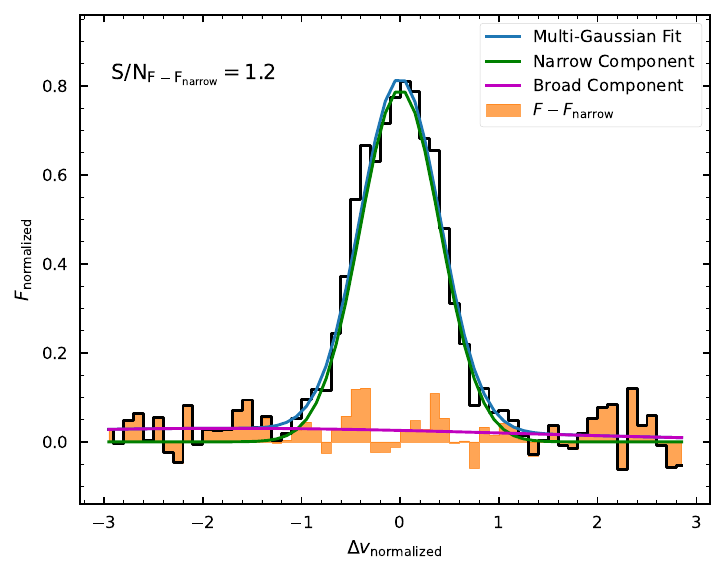}
  \centering (b) non-BAL stack (velocity-normalized)
\end{minipage}
\caption{Velocity-normalized stacking of [C\,{\sc ii}] spectra for the BAL (panel a) and non-BAL (panel b) quasar samples. Each individual profile has been normalized by its FWHM before stacking. A broad component remains visible in the BAL stack, while the non-BAL stack shows no significant excess. \added{This is consistent with the BAL broad residual not being driven solely by differences in intrinsic narrow-line widths.}}
\label{fig:vnorm-stack}
\end{figure*}

\subsection{Intrinsic Properties of Quasars and Sample Size}
Here, we discuss potential systematics and caveats that could influence our results. Previous studies have found a positive correlation between the incidence of [C\,{\sc ii}] outflows and quasar bolometric luminosity, suggesting that more luminous quasars are more likely to drive large-scale outflows \citep{bischetti_widespread_2019, maiolino_evidence_2012}. Since direct measurements of bolometric luminosity are not available for our sample, we use $M_{1450}$ as a proxy, assuming a typical bolometric correction for high-redshift quasars. Figure~\ref{fig:M1450} shows the distribution of our quasar sample in the $M_{1450}$ versus redshift plane. Our sample spans a narrow redshift range, and both BAL and non-BAL quasars exhibit similar distributions in $M_{1450}$. This similarity reduces the likelihood that luminosity-related biases are responsible for the observed differences in [C\,{\sc ii}] profiles between the two samples.

To test whether the observed broad [C\,{\sc ii}] wing is driven by brighter quasars that preferentially exhibit BAL features, we divide the sample into two subgroups based on $M_{1450}$: quasars brighter than $M_{1450} = -27$ and those fainter than $M_{1450} = -27$. We stack the [C\,{\sc ii}] spectra separately for each subgroup and find no significant (greater than 2$\sigma$) detection of a broad residual or a systematically blueshifted [C\,{\sc ii}] line in either stack. This result suggests that quasar brightness is unlikely to be the primary factor driving the broad [C\,{\sc ii}] outflows in BAL quasars, at least within our sample.

Due to the absence of spectral coverage for Mg\,{\sc ii} or rest-frame optical emission lines, we cannot reliably constrain black hole masses or Eddington ratios for our sample. However, previous studies have shown that the occurrence of BAL features is not strongly correlated with these properties \citep{bischetti_fraction_2023}, mitigating potential concerns about their influence on our findings.

A significant remaining caveat is the lack of precise constraints on the host galaxy masses, as differences in host properties could affect the interpretation of our results. We note that the mean Gaussian FWHMs of individual quasars are similar between the BAL ($4.3\times10^2\,\rm km\,s^{-1}$) and non-BAL ($4.6\times10^2\,\rm km\,s^{-1}$) samples, suggesting no strong difference in dynamical masses. Additionally, without Mg\,{\sc ii} coverage, we cannot determine whether any BAL quasars in our sample belong to the low-ionization BAL (LoBAL) subclass, which may be intrinsically distinct \citep{bischetti_multiphase_2024}. Therefore, we cannot assess whether intrinsic physical differences between BAL and non-BAL host galaxies contribute to the observed differences in [C\,{\sc ii}] profiles.

Nevertheless, we conduct an additional robustness check to verify that the broad [C\,{\sc ii}] component detected in the BAL quasar stack is not simply due to statistical fluctuations arising from the small sample size or potential systematics from data reduction. We randomly select five quasars from the non-BAL group, stack their spectra, and measure the significance of any broad [C\,{\sc ii}] wing. Repeating this process 1,000 times, we find that the cumulative probability of randomly obtaining a broad wing with a significance equal to or greater than S/N = 4.4 is less than 2\% (Figure~\ref{fig:small-sample}). \added{This disfavors the possibility that the full-stack BAL broad residual arises purely from chance alignments of noise in a five-object stack.} \added{However, the evidence remains statistical, and the conservative clean-stack analysis yields only a hint-level broad residual.} \added{We therefore treat the broad [C\,{\sc ii}] wing as suggestive and avoid drawing a definitive causal connection based on the present sample.}

\added{While it would be valuable to expand the analysis to additional quasar samples in the literature (e.g., WISSH at $z<5$, \citealp{bischetti_wissh_2018}; or XQR-30 at $z\gtrsim 6$, M.\ Neeleman et al. in prep), combining datasets from multiple ALMA programs would introduce heterogeneity in redshift, sensitivity, spectral resolution, and angular resolution, which can complicate continuum treatment and bias stacked line profiles. We therefore focus our primary analysis on a homogeneous ALMA sample at $z\sim5.5$ to minimize these systematics and enable a controlled comparison between BAL and non-BAL subsamples. Complementary high-redshift stacking analyses are being presented in S.\ Shanbhog et al.\ (in prep.) using the XQR-30 quasars at $z\gtrsim6$, who also find a blueshifted broad component in stacked [C\,{\sc ii}] profiles.}

~

\subsection{Robustness of Stacking Methods and Interpretation}

We note that \citet{novak_no_2020} reported no clear evidence of [C\,{\sc ii}] outflows in quasar host galaxies at $z > 6$, although their high-resolution observations were not optimized for detecting outflows on kiloparsec scales (see also \citealp{decarli_alma_2018,sawamura_no_2025}). In their analysis, \citet{novak_no_2020} employed UV-plane stacking (in addition to image-plane stacking) to enhance sensitivity and control for systematics arising from variable observing times and array configurations. In contrast, our dataset was collected using a uniform observational setup and consistent ALMA configurations, which already minimize these systematic concerns. Therefore, UV-plane stacking was unnecessary for our study. Nevertheless, as a consistency check, we performed image-plane stacking, as shown in Figure~\ref{fig:cube-stack}, confirming that our main conclusions remain unaffected by this stacking method. When directly stacking image cubes, the BAL sample still exhibits a broad [C\,{\sc ii}] component with a significance of $4.48\sigma$, while the non-BAL stack remains below the $2\sigma$ detection threshold.

We caution that the method of continuum subtraction might still influence the visibility of the broad [C\,{\sc ii}] wing. \added{However, as described in Section~\ref{sec:data}, we subtract the dust continuum in the $uv$ domain using line-free channels prior to imaging and spectral extraction.} \added{We verified that adopting an alternative continuum subtraction using a zeroth-order polynomial restricted to the emission-line spectral window in the $uv$ domain does not change our results.} Additionally, based on random stacking tests of non-BAL quasars (Figure~\ref{fig:small-sample}), our continuum subtraction method is unlikely to introduce the observed blueshifted broad wing systematically. \added{Any residual continuum offset would be approximately velocity-independent over the fitted window and would not naturally produce a localized, blueshifted broad residual relative to the narrow-line model.} \added{We also verified that allowing an additional constant offset term in the stacked-line fitting does not change the inferred broad-component S/N or centroid.} Throughout data reduction, we had no prior expectation of a broad [C\,{\sc ii}] component and consistently followed the same procedures across the sample, further reducing the risk of systematic biases.

To examine the influence of individual quasars on our results, we conducted multiple tests:

\begin{enumerate}
    \item The quasar J1513 exhibits a double-peaked [C\,{\sc ii}] profile, indicative of a rotating disk. Excluding this quasar does not affect our conclusions regarding the non-detection of a blueshifted broad component in the non-BAL stack.
    

    \item Some ALMA spectra exhibit incomplete spectral coverage of the [C\,{\sc ii}] line (BAL: J0012; non-BAL: J1006 and J1016). \added{Adopting the conservative clean-stack selection (excluding these quasars), the BAL broad residual decreases to a hint-level feature with ${\rm S/N}=2.85$, while the non-BAL stack remains consistent with no obvious broad residual (${ \rm S/N}=1.93$).} \added{Because the clean BAL stack removes J0012, this provides a direct check that partial coverage in a single spectrum does not by itself drive the full-stack wing.} \added{J0012 contributes noticeably in the full stack, and its missing red-side channels could in principle bias the fitted wing strength.} \added{However, J0012 also shows an intrinsically broad [C\,{\sc ii}] profile in the individual spectrum (Figure~\ref{fig:cii}), and the BAL jackknife tests (Figure~\ref{fig:jackknife}) show that the broad residual remains visible when any other BAL quasar is removed, with the lowest S/N occurring when excluding J0012.} \added{We therefore treat the full-stack wing amplitude and the associated energetics as approximate upper limits.}
    
\item To directly test the influence of individual sources, we performed jackknife tests by excluding one BAL quasar at a time and re-stacking the remaining four. The results are shown in Figure~\ref{fig:jackknife}. In all five cases, the broad component remains \added{visible}, with measured significance values \added{greater than} $2.8\sigma$\added{, corresponding to hint-level features under our adopted S/N classification}. The lowest value ($2.85\sigma$) occurs when J0012 is excluded, consistent with its individually broad profile but incomplete spectral coverage. These tests \added{suggest that the broad [C\,{\sc ii}] wing is not solely driven by a single source, although its significance is reduced when excluding J0012.}
    
    \item To investigate the potential impact of host galaxy kinematics, we separated quasars into high- and low-FWHM groups (divided at 450 km\,s$^{-1}$, close to the mean or median FWHM). Neither group individually yields $>3\sigma$ detection of a broad component (low-FWHM: $2.61\sigma$, high-FWHM: $2.17\sigma$), indicating that host galaxy kinematics alone do not explain the observed differences. The non-BAL group shows a broader range of line widths, yet does not show evidence for a broad wing, further arguing against FWHM-driven effects.
    
\item We tested normalization of spectra by their FWHM and amplitude, following procedures adopted by previous studies such as \citet{novak_no_2020}. This normalization is intended to account for variation in intrinsic [C\,{\sc ii}] line width across the sample, and mitigates biases due to particularly broad or narrow systems. The resulting velocity-normalized stacks are shown in Figure~\ref{fig:vnorm-stack}. While this normalization slightly reduces the broad-component significance in the BAL group to $2.8\sigma$, the broad wing remains visible. The non-BAL group continues to show no \added{obvious} broad component ($1.0\sigma$). \added{This is consistent with the BAL broad residual not being solely driven by a subset of intrinsically broader narrow-line systems.} However, we caution that this normalization might dilute weak broad wings, especially if the wing and the line core have different physical origins. Thus, the lower significance in this test may reflect an underestimation of true broad wing strength.
    
    \item For ESI spectra, due to limited wavelength coverage and uncertainties from telluric absorption, weak BAL features might be ambiguous. To address this, we separately stacked only the most robust BAL quasars (${\rm BI}_0 > 2500$ km\,s$^{-1}$; J2317, J2325) against definitive non-BAL quasars (${\rm BI}_0 = 0$; J0120, J0231, J0306, J1500, J1513, J1614), normalizing by both amplitude and FWHM. We still detect a broad component at $2.38\sigma$ for the robust BAL subgroup versus $1.15\sigma$ for definitive non-BAL quasars, reinforcing the robustness of our conclusion despite smaller samples.
\end{enumerate}

Collectively, these tests \added{show that the presence of a broad residual in the full-stack BAL spectrum is not an obvious artifact of continuum subtraction, normalization choices, or a single-object anomaly.} \added{However, the inferred significance depends on sample definition, and the conservative clean-stack analysis yields only a hint-level broad residual.} We emphasize that our BAL sample, while modest, already represents the largest high-redshift BAL quasar sample with uniform [C\,{\sc ii}] observations currently available, built up over the past five years. \added{Larger and more uniform samples, together with spatially resolved follow-up, will be required to establish the prevalence and physical origin of [C\,{\sc ii}] wings and to test any connection between nuclear and galactic-scale outflows during the reionization epoch.}

\subsection{Individual Quasar: J0056+2241}

\begin{figure}[!ht]
    \centering
    \includegraphics[width=1\linewidth]{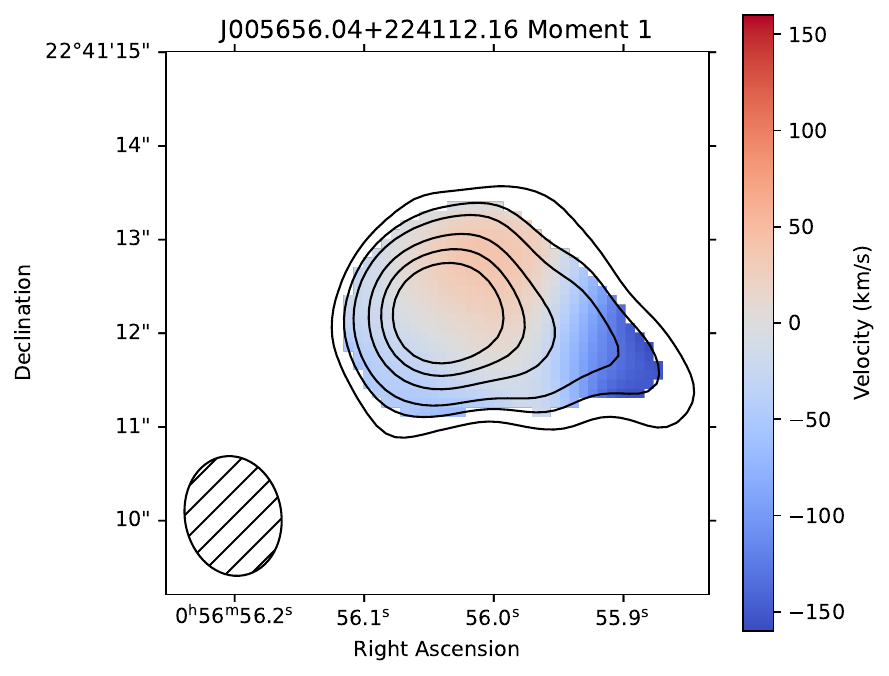}
    \caption{The moment 1 (velocity) map of J0056+2241 relative to its systemic redshift of $z=5.5218$. The contours represent significance levels of (2$\sigma$, 3$\sigma$, 4$\sigma$, 6$\sigma$, 8$\sigma$, 10$\sigma$) from the moment 0 map. The map reveals blueshifted emission within the extended [C\,{\sc ii}] structure, although the velocity remains within the FWHM of its emission line profile, as shown in Figure \ref{fig:cii}.}
    \label{fig:J0056}
\end{figure}

Among our quasar sample, the non-BAL quasar J0056+2241 exhibits an exceptionally extended morphology in its [C\,{\sc ii}] emission map, with a tail extending over $2''.5$ (corresponding to $\sim 15$ kpc) toward the southwest from the quasar center. The moment 1 map (Figure \ref{fig:J0056}) reveals blueshifted emission within the extended [C\,{\sc ii}] structure, with a velocity offset of $\sim - 100$--$150\ \rm km\,s^{-1}$, although this velocity remains within the width of its emission line profile (FWHM = 570 km s$^{-1}$; Figure \ref{fig:cii}). This large-scale structure may indicate a significant outflow of cold gas on galactic scales, possibly driven by strong AGN activity. However, such an extended structure could also result from other processes, such as a merger \citep{banados_z_2019,izumi_subaru_2021-1}.

Interestingly, despite this potential morphological evidence of outflows, the ESI spectrum of J0056+2241 does not show strong BAL features. However, there is a notable flux discontinuity redward of the Ly$\alpha$ + [N\,{\sc v}] emission complex. Additionally, the Ly$\alpha$ line itself is associated with strong absorption features, as also noted in \citet{zhu_probing_2023}. These absorbers may indicate the presence of intervening gas clouds or infalling material along the line of sight.

One potential explanation for the lack of BAL features is that the outflows in J0056+2241 may not be well-aligned with our line of sight. Indeed, we do not see strong blueshifted [C\,{\sc ii}] emission, suggesting that the [C\,{\sc ii}] outflow is also not aligned along our line of sight. In this scenario, the outflow may be oriented at an angle where high-velocity winds do not produce strong absorption features in the rest-frame UV spectrum. Alternatively, the outflow may be dominated by dense, neutral gas phases that are less effective at producing BAL signatures in high-ionization species such as C\,{\sc iv} and Si\,{\sc iv}. 

Another possibility is a difference in timescales between nuclear and galactic-scale outflows. If the large-scale outflow observed in [C\,{\sc ii}] is a remnant of past AGN activity, the current nuclear wind may no longer be strong enough to produce detectable BAL features in the UV. Alternatively, J0056+2241 may be in an early stage of outflow development, where neutral gas has already been accelerated to large scales, but the ionized phase has not yet fully developed. Further spatially resolved spectroscopic observations, particularly in ionized and molecular gas tracers, could help disentangle the structure and dynamics of the outflow in J0056+2241, providing additional insight into the feedback mechanisms in this system.

Finally, as a sanity check, including J0056+2241 in our non-BAL quasar stack increases the significance of the broad component only from 1.72$\sigma$ to 1.77$\sigma$, which is slightly higher than when excluding J0056+2241. This suggests that its extended [C\,{\sc ii}] emission may contribute modestly to the overall broad wing detected in the stacked profile.

\section{Summary}\label{sec:summary}

We present evidence for a potential connection between nuclear winds and large-scale cold gas outflows in high-redshift ($z \sim 5.5$) quasars, based on a stacking analysis of ALMA [C\,{\sc ii}] 158 $\mu$m emission line profiles in BAL and non-BAL sub-samples. Our main findings are:

\begin{enumerate}
    \item The stacked [C\,{\sc ii}] spectrum for BAL quasars reveals a \added{blueshifted} broad component, with a velocity offset of $\Delta v_{\rm b} = -2.1\times10^2 \, \rm km\,s^{-1}$ and a FWHM of $1.18\times10^3 \, \rm km\,s^{-1}$, consistent with large-scale cold gas outflows, \added{although the conservative clean-stack analysis yields a hint-level broad residual}. The non-BAL stack shows no statistically significant broad component, \added{and the clean non-BAL stack shows no obvious broad residual}. While this difference may reflect a physical link between BAL winds and cold gas outflows, alternative explanations such as galaxy orientation or differing host properties cannot be excluded.

    \item We estimate a coupling efficiency of $\eta = 0.025_{-0.022}^{+0.23}$ between the kinetic power of BAL outflows and that of the [C\,{\sc ii}]-traced outflows. This suggests that BAL winds may transfer a non-negligible fraction of their energy into neutral gas on kiloparsec scales, \added{but the inferred energetics should be regarded as illustrative and as approximate upper limits given the hint-level clean-stack result and other uncertainties}. However, this estimate assumes that the [C\,{\sc ii}] broad component is AGN-driven; contributions from star formation-driven winds would reduce the inferred efficiency.

    \item The non-BAL quasar J0056+2241 displays an extended [C\,{\sc ii}] tail spanning $\sim$15 kpc with blueshifted emission, suggestive of an outflow or merger-induced disturbance. Despite lacking strong BAL features, its morphology implies that orientation, feedback history, or merging activity may play a role in shaping large-scale gas dynamics.
\end{enumerate}

Taken together, our results support a scenario in which nuclear outflows may contribute to cold gas removal in quasar host galaxies during the reionization epoch, \added{but the conservative clean-stack analysis indicates that the current evidence is at the hint level.}
\added{We caution that the interpretation is limited by sample size, uncertainties in star formation contributions, and potential orientation biases.} Further spatially resolved, multi-phase observations will be essential to robustly constrain the physical origin of [C\,{\sc ii}] wings and the role of BAL winds in early feedback processes. If confirmed with a larger sample, given the high BAL incidence at $z > 5$ \citep{bischetti_suppression_2022}, such winds may be a key component of AGN-host co-evolution at the end of reionization.

\section*{Acknowledgments}
\added{We thank the anonymous reviewer for providing very helpful feedback and suggestions.}
YZ, MJR, YS, GHR, and ZJ acknowledge support from the NIRCam Science Team contract to the University of Arizona, NAS5-02105. This work was supported by the NSF through award SOSPADA-029 from the NRAO. YZ and GDB were supported by the NSF through grant AST-1751404. 
LCH was supported by the National Science Foundation of China (12233001) and the National Key R\&D Program of China (2022YFF0503401). FY is supported by the Natural Science Foundation of China (grants 12133008, 12192220, 12192223, and 12361161601), the China Manned Space Program through its Space Application System, and the National Key R\&D Program of China No. 2023YFB3002502. TTT has been supported by the Japan Society for the Promotion of Science (JSPS) Grants-in-Aid for Scientific Research (24H00247). 
This work has also been supported in part by the Collaboration Funding of the Institute of Statistical Mathematics ``Machine-Learning-Based Cosmogony: From Structure Formation to Galaxy Evolution''. This manuscript benefited from grammar checking and proofreading using ChatGPT \citep{openai_chatgpt_2023}.
This paper makes use of the following ALMA data: ADS/JAO.ALMA\allowbreak\#2022.1.00662.S. ALMA is a partnership of ESO (representing its member states), NSF (USA) and NINS (Japan), together with NRC (Canada), MOST and ASIAA (Taiwan), and KASI (Republic of Korea), in cooperation with the Republic of Chile. The Joint ALMA Observatory is operated by ESO, AUI/NRAO and NAOJ. The National Radio Astronomy Observatory and Green Bank Observatory are facilities of the U.S. National Science Foundation operated under cooperative agreement by Associated Universities, Inc.
Some of the data presented herein were obtained at Keck Observatory, which is a private 501(c)3 non-profit organization operated as a scientific partnership among the California Institute of Technology, the University of California, and the National Aeronautics and Space Administration. The Observatory was made possible by the generous financial support of the W. M. Keck Foundation. The authors wish to recognize and acknowledge the very significant cultural role and reverence that the summit of Maunakea has always had within the Native Hawaiian community. We are most fortunate to have the opportunity to conduct observations from this mountain. This research has made use of the Keck Observatory Archive (KOA), which is operated by the W. M. Keck Observatory and the NASA Exoplanet Science Institute (NExScI), under contract with the National Aeronautics and Space Administration.

\vspace{5mm}
\facilities{ALMA; Keck:II (ESI)}

\software{
    {\tt Astropy} \citep{astropy_collaboration_astropy_2013, astropy_collaboration_astropy_2018, astropy_collaboration_astropy_2022},
    {\tt CARTA} \citep{comrie_carta_2021},
    {\tt CASA} \citep{mcmullin_casa_2007,casa_team_casa_2022},
    {\tt GDL} \citep{coulais_status_2010}, 
    {\tt Matplotlib} \citep{hunter_matplotlib_2007},
    {\tt NumPy} \citep{van_der_walt_numpy_2011},
}

\section*{Author Contributions}
YZ led the observations and data reduction for the ALMA and ESI programs, performed the initial analysis, and wrote the first draft. MJR, LCH, YS, GHR, and FY contributed to the initial discussions and interpretation of the results. TJLCB, GDB, and JY assisted with observations and sample selection. EB and MB contributed to the analysis methodology and statistical interpretation. All authors contributed to the presentation and discussions.

\appendix

\renewcommand{\thefigure}{A\arabic{figure}}
\setcounter{figure}{0}

\section{The Quasar Sample}

\begin{figure*}[!ht]
    \centering
    \includegraphics[width=0.9\linewidth]{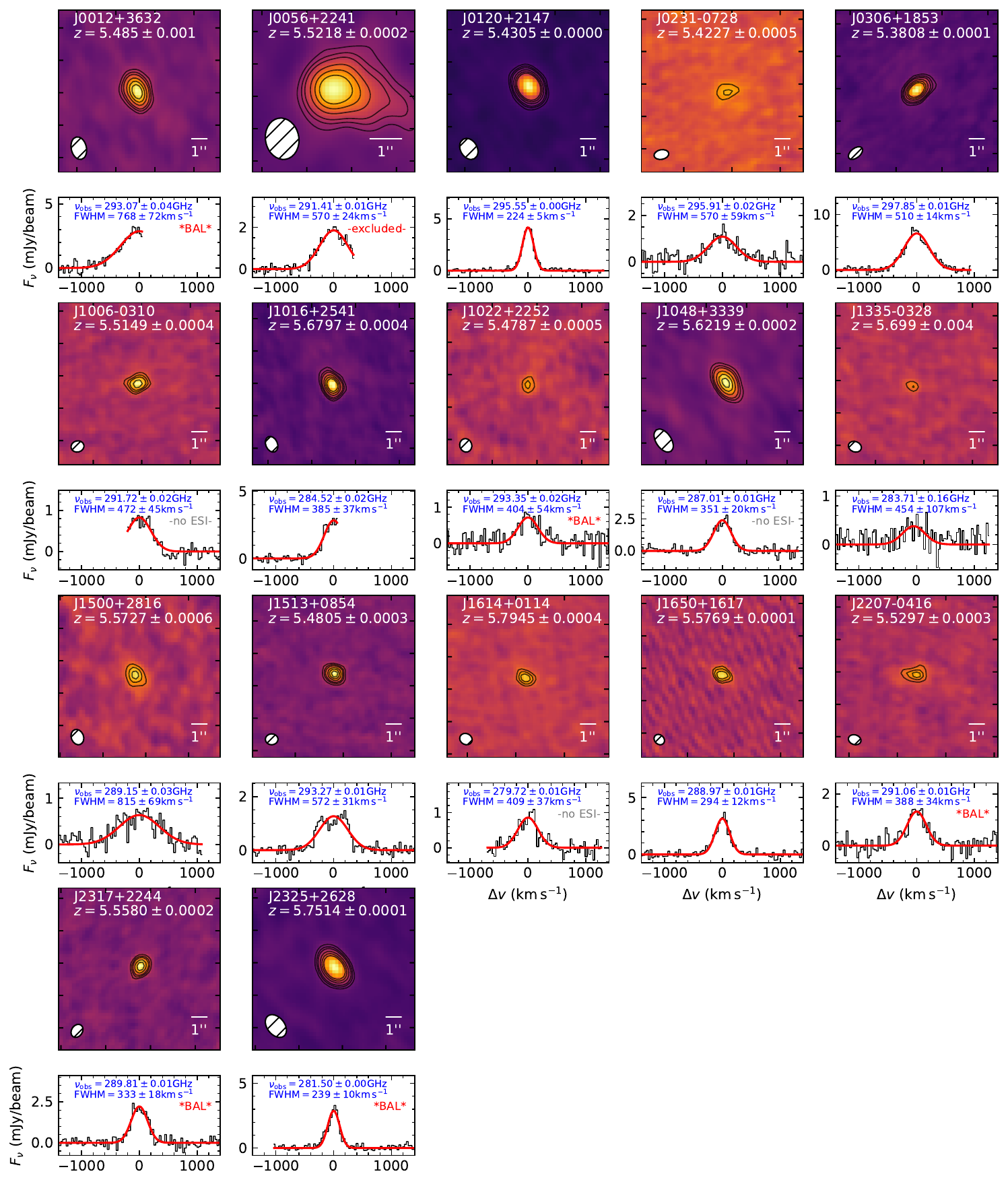}
    \caption{[C\,{\sc ii}] 158$\mu$m emission maps and spectra of the quasar sample observed with ALMA. Contours show (2$\sigma$, 3$\sigma$, 4$\sigma$, 6$\sigma$, 8$\sigma$, 10$\sigma$) levels. The synthesized beam is shown in the lower-left corner of each map. Measured redshifts are labeled for each quasar and red curves show the best Gaussian fits. Observed frequency and FWHM of the [C\,{\sc ii}] emission are also provided for reference. Quasars with significant BAL features are labeled, while those without ESI spectra are marked as ``no ESI''. Since J0056 exhibits very extended [C\,{\sc ii}] emission in the image, it has been excluded from the stacking analysis.}
    \label{fig:cii}
\end{figure*}

\begin{figure*}[!ht]
    \centering
    \includegraphics[width=0.95\linewidth]{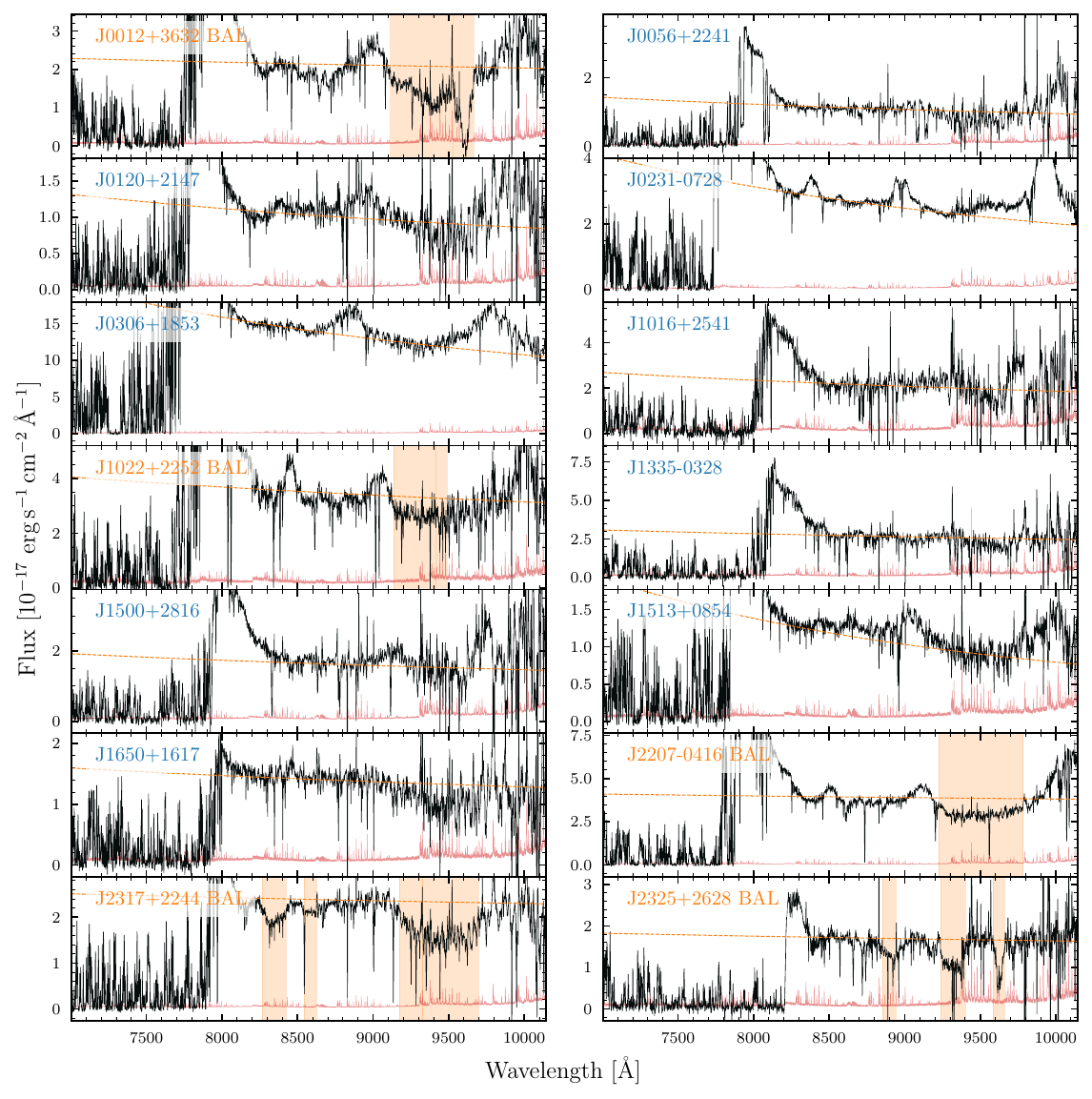}
    \caption{Keck/ESI rest-frame UV spectra of the quasar sample. Black and red colors plot the flux and corresponding error array, respectively. BAL quasars (${\rm BI}_0 > 1000\ \rm km\, s^{-1}$) are labeled with orange text, and regions shaded in orange highlight prominent BAL absorption troughs. The dashed curves indicate the power-law continuum fits used to normalize the spectra. 
    \added{The spectra are available in CSV format at \dataset[doi:10.5281/zenodo.18677316]{\doi{10.5281/zenodo.18677316}}}.
    }
    \label{fig:ESI}
\end{figure*}

Table \ref{tab:QSOlist} lists the properties of the quasars used in this work. Figure \ref{fig:cii} shows the ALMA observations and Figure \ref{fig:ESI} presents the Keck/ESI spectra. \added{The reduced Keck/ESI spectra used for BAL classification are available at \dataset[doi: 10.5281/zenodo.18677316]{\doi{10.5281/zenodo.18677316}}.}

\begin{deluxetable*}{llcccccc}
    \tablenum{1}
    \tablecaption{Quasar Sample Used in This Work}
    \tablehead{
        \colhead{\hspace{0.258cm}No.}\hspace{0.258cm} & 
        \colhead{\hspace{0.258cm}Quasar}\hspace{0.258cm} & 
        \colhead{\hspace{0.258cm} RA (J2000)}\hspace{0.258cm} & 
        \colhead{\hspace{0.258cm} Dec (J2000)}\hspace{0.258cm} & 
        \colhead{\hspace{0.258cm}$z_{\rm [CII]}$}\hspace{0.258cm} & 
        \colhead{\hspace{0.258cm}$M_{1450}$}\hspace{0.258cm} & 
        \colhead{\hspace{0.258cm}$\rm BI_0\,(km\,s^{-1})$} \hspace{0.258cm} & 
        \colhead{\hspace{0.258cm}ESI}\hspace{0.258cm} 
    }
    \decimalcolnumbers
    \startdata
1 & J0012$+$3632 & 00:12:32.88 & $+$36:32:16.10 & $5.485 \pm 0.001$ & -27.2 & \textbf{5650.9} & 2021-10 \\
2 & J0056$+$2241$^{\rm a}$ & 00:56:56.04 & $+$22:41:12.16 & $5.5218 \pm 0.0002$ & -26.8 & 0.0 & 2022-09 \\
3 & J0120$+$2147 & 01:20:53.92 & $+$21:47:06.20 & $5.4305 \pm 0.0001$ & -26.5 & 0.0 & 2023-09 \\
4 & J0231$-$0728 & 02:31:37.64 & $-$07:28:54.44 & $5.4227 \pm 0.0005$ & -26.6 & 0.0 & 2013-01 \\
5 & J0306$+$1853 & 03:06:42.51 & $+$18:53:15.85 & $5.3808 \pm 0.0001$ & -28.9 & 0.0 & 2021-10 \\
6 & J1006$-$0310$^{\rm b}$ & 10:06:14.61 & $-$03:10:30.49 & $5.5149 \pm 0.0004$ & -27.0 & 0.0 & - \\
7 & J1016$+$2541 & 10:16:37.70 & $+$25:41:31.98 & $5.6797 \pm 0.0004$ & -27.7 & 471.8 & 2024-06 \\
8 & J1022$+$2252 & 10:22:10.04 & $+$22:52:25.44 & $5.4787 \pm 0.0005$ & -27.3 & \textbf{1501.9} & 2016-03 \\
9 & J1048$+$3339$^{\rm c}$ & 10:48:36.72 & $+$33:39:47.66 & $5.6219 \pm 0.0002$ & -27.0 & 0.0 & - \\
10 & J1335$-$0328 & 13:35:56.23 & $-$03:28:38.20 & $5.699 \pm 0.004$ & -27.8 & 304.8 & 2024-05 \\
11 & J1500$+$2816 & 15:00:36.83 & $+$28:16:03.03 & $5.5727 \pm 0.0006$ & -27.6 & 0.0 & 2024-06 \\
12 & J1513$+$0854 & 15:13:39.64 & $+$08:54:06.58 & $5.4805 \pm 0.0003$ & -26.8 & 0.0 & 2021-05 \\
13 & J1614$+$0114$^{\rm c}$ & 16:14:35.35 & $+$01:14:44.79 & $5.7945 \pm 0.0004$ & -26.9 & 0.0 & - \\
14 & J1650$+$1617 & 16:50:42.25 & $+$16:17:21.50 & $5.5769 \pm 0.0001$ & -27.2 & 477.7 & 2021-05 \\
15 & J2207$-$0416 & 22:07:10.12 & $-$04:16:56.28 & $5.5297 \pm 0.0003$ & -27.8 & \textbf{2447.0} & 2021-10 \\
16 & J2317$+$2244 & 23:17:38.25 & $+$22:44:09.63 & $5.558 \pm 0.0002$ & -27.4 & \textbf{4329.0} & 2022-09 \\
17 & J2325$+$2628 & 23:25:14.24 & $+$26:28:47.61 & $5.7514 \pm 0.0001$ & -27.0 & \textbf{2676.2} & 2021-10 \\
    \enddata
    \tablecomments{Columns: (1) Index; (2) quasar name; (3) and (4) quasar coordinates (J2000); (5) [C\,{\sc ii}] redshift; (6) absolute magnitude at rest-frame 1450 \AA, based on quasar luminosities from \citet{yang_discovery_2017, yang_filling_2019, zhu_probing_2023}; (7) balnicity, where strong BAL quasars (${\rm BI}_0 > 1000$ km s$^{-1}$; see text for details) are marked in bold; (8) date of the most recent Keck/ESI observation.
    \\
    a: outflow signature shown in the [CII] image.\\
    b: see \citet{yang_discovery_2017} for spectra.\\
    c: see \citet{yang_filling_2019} for spectra.
    }
    \label{tab:QSOlist}
\end{deluxetable*}

\pagebreak

\bibliographystyle{aasjournalv7}

\end{document}